\documentclass[a4paper,10pt]{article}

\usepackage[usenames,dvipsnames,svgnames,table]{xcolor}

\usepackage[pdftitle={Fishing out collective memory of migratory schools}, pdfauthor={Giancarlo De Luca, Patrizio Mariani, Brian MacKenzie and Matteo Marsili}, pdfsubject={collective intelligence | network dynamics | schooling | bluefin tuna}, pdfcreator={created with LaTeX and pdfLaTeX}]{hyperref}

\usepackage{abstract}
\usepackage[utf8x]{inputenc}
\usepackage{setspace}
\usepackage[english]{babel}
\usepackage[T1]{fontenc}
\usepackage{amsmath,amssymb}
\usepackage{mathptmx}
\usepackage{authblk}

\usepackage{graphicx}
\usepackage{bm}
\usepackage[normalem]{ulem}

\newcommand{\bvec}[1]{\mathbf{#1} }
\newcommand{\ee}{\mathrm{e}}

\newcommand{\fn}[1]{\mathrm{#1}}

\DeclareMathOperator{\Prob}{\mathrm{Prob}}
\newcommand{\keywords}[1]{\smallskip \noindent \textbf{Keywords:}  #1}

\usepackage{enumerate}

\usepackage{mathrsfs}
\usepackage{bm}
\usepackage[normalem]{ulem}
\usepackage{cite}
\newcommand{\mat}[1]{\mathsf{#1}}
\newcommand{\sset}[1]{\mathscr{#1}}

\DeclareGraphicsExtensions{.eps, .pdf, .ps, .jpg}

\title{Fishing out collective memory of migratory schools}

\author[1]{Giancarlo De Luca\thanks{ Now at: IPLESP---Institut Pierre Louis d'Epid\'{e}miologie et de 
Sant\'{e} Publique, INSERM  \& UPMC UMR-S 1136, , Paris, 
France. Email: 
giancarlo.de-luca@inserm.fr}}
\author[2]{Patrizio Mariani}
\author[2,4]{Brian R. MacKenzie}
\author[5]{Matteo Marsili}

\affil[1]{\textsf{\small SISSA -- International School 
for Advanced Studies, Trieste, Italy}}

\affil[2]{\textsf{\small Centre for Ocean Life,
National Institute for Aquatic Resources, Technical University of Denmark, 
Charlottenlund, Denmark} }
\affil[3]{\textsf{\small Center for Macroecology, 
Evolution and Climate, National Institute for Aquatic Resources, Technical University of Denmark
Charlottenlund, Denmark}}
\affil[4]{\textsf{\small The Abdus Salam International 
Centre For Theoretical Physics, Trieste, Italy}}

\begin{document}

\maketitle

\begin{abstract}
Animals form groups for many reasons but there are costs and benefit 
associated with group formation. One of the benefits is collective memory. 
In groups on the move, social interactions play a crucial role 
in the cohesion and the ability to make consensus decisions. 
When migrating from spawning to feeding areas fish schools need to retain a 
collective memory of the destination site over thousand of kilometres and 
changes in group formation or individual preference can produce sudden changes 
in migration pathways.
We propose a modelling framework, based on stochastic adaptive networks, that can reproduce this collective behaviour. 
We assume that three factors control group formation and school migration behaviour: the intensity of social 
interaction, 
the relative number of informed individuals and the strength of preference that informed individuals have for a 
particular 
migration area. 
We treat these factors independently and relate the individuals' preferences to the 
experience and memory for certain migration sites.
We demonstrate that removal of knowledgeable individuals or alteration of individual 
preference can produce rapid changes in group formation and collective behaviour.
For example, intensive fishing targeting the migratory species and also their preferred prey 
can reduce both terms to a point at which migration to the destination sites
is suddenly stopped. The conceptual approaches represented by our modelling framework may therefore be able to explain 
large-scale changes in fish migration and spatial distribution.
\end{abstract}


\section{Introduction}

Grouping behaviour is a widespread phenomenon in animal ecology and is thought to be an emerging 
property of the self-organisation of individual organisms~\cite{Sumpter2010}. 
While living in groups, social animals benefit from several advantages among which is a more 
efficient capacity in problem solving~\cite{krause2002living, Krause2010,Surowiecki2005}. 
Of particular interest is the ability of the group to make collective decisions also when it is
composed of individuals with contrasting preferences and information~\cite{Lachlan1998,Conradt2011}. 
How groups reach a consensus decision has recently received 
 much attention~\cite{sumpter2008consensus,Couzin2011,Ward2011,Miller2013} and several mechanisms to pool 
information in the group have been proposed~\cite{Conradt2011,Sumpter2010}. 

Often no obvious reason can be 
adduced to explain the social behaviour of certain species except the fact that those groups are more efficient than 
single individuals in retrieving information 
from the environment~\cite{Conradt2011a,Ward2011,berdahl2013emergent}. 
For groups on the move, such as fish schooling, bird flocking or mammal herding, it 
has been shown that information transfer and social interactions are important factors of group 
cohesion and can promote the ability of 
making consensus decisions~\cite{Brown2003,Couzin2005,Miller2013}.

An example of such a collective decision making problem is the structure of migration routes in 
some fish species. Migration between widely separated but geographically stable locations of 
spawning and feeding sites raises several questions about how these animals manage 
to learn and \emph{remember} the migration route between feeding and spawning sites. 
Where is the information on the path stored? 
How is it retrieved, shared and elaborated by a migrating group?
 Are these tasks performed significantly better by the group with respect to the individuals?   
Shedding light on the functioning of these mechanisms is a fundamental issue in ecology but may 
also be relevant to fields such as sociology and economy where it is common to deal with large systems 
of competitive agents that share information~\cite{Conradt2011a,Surowiecki2005}.
We hypothesize that \emph{collective memory} might play an important role in 
the migration process of fish populations~\cite{Brown2003} 
and model its effects on schooling behaviour and migration efficiency.
We tackle these questions by assuming that individuals have different amounts of 
information 
about migration routes and that only a fraction of them possesses some information, whereas the rest only exhibit a 
social behaviour. Those assumptions are consistent with numerical simulations 
of the evolution of leader and social traits in migratory populations~\cite{Guttal2010,Guttal2011} but 
are introduced in our model in a different way. 
In fact previous approaches mainly fall in a class of agent based models with 
spatial interaction~\cite{Katz2011,Vicsek2012,Sumpter2010,Chate2008,Guttal2010} where ``social'' 
individuals tend to align and to follow the individuals that are nearby, in a finite spatial range. 
This reproduces a realistic dynamics, but it gives little 
insight on the mechanisms by which the collective behaviour emerges from individual interactions. Indeed due to their 
complexity, spatial dynamics models 
can only be studied with extensive numerical simulations.

\begin{figure}[!htbp]
\centering
\includegraphics{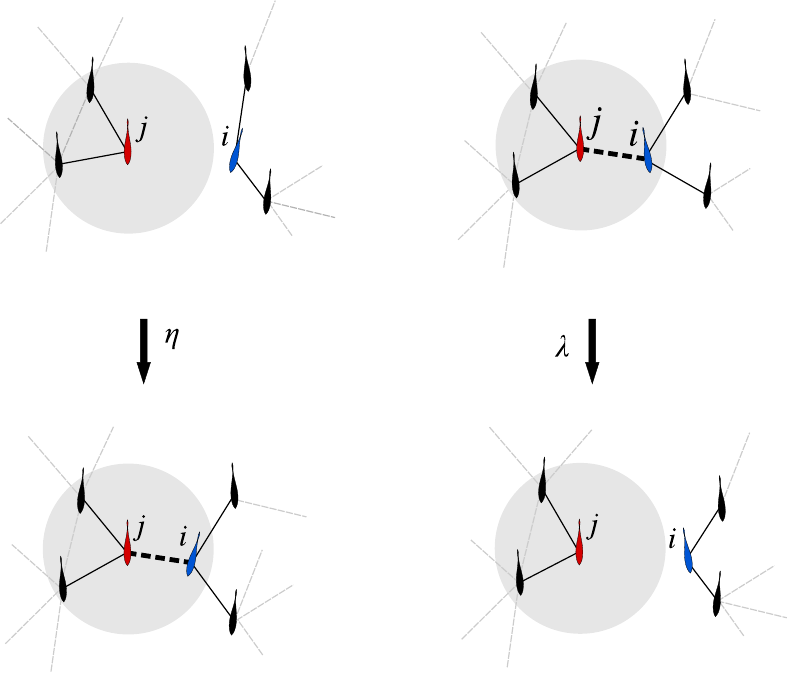}
\caption{Connection between link creation and destruction process and real space 
models where $\eta$ is the link creation rate and $\lambda$ the link 
destruction rate. }\label{fig:net_up_proc}
\end{figure}

Here, instead, we take {a stochastic adaptive network approach. Network approaches have already been successfully 
applied to 
address collective behaviour in animal  groups
~\cite{Huepe2011,Gross2008,Couzin2011}: adaptive network models provide, in fact, 
a simpler mathematical structure which can be analysed more easily than real 
space models (i.e. without relying on simulations). In all these models, as in ours, spatial dynamics is implicitly 
taken into account through link 
creation and destruction processes: changes in the neighbourhood of the 
individuals due to spatial dynamics are reproduced by link dynamics between nodes (see Figure~\ref{fig:net_up_proc}).

}

Capitalizing on previous models~\cite{Marsili2010,Ehrhardt2006}, { we build a model} introducing the key ingredient of 
memory for preferred route directions in a fraction of the individuals (the informed ones). This is introduced as an 
\emph{a priori} bias for a particular route in the choice behaviour of the informed individuals, that is based on their 
experience in that particular habitat: their \emph{memory}.
Therefore, the collective choice of the route direction, is a function of individual and social processes. 
We are able to find an exact solution for the model that provides a clear picture of how information is 
elaborated, stored and shared in the group and allows us to describe an observed 
switch of migratory 
path in fish populations as a result of a loss of group level information.  

\section{Theoretical framework}
Most studies about swarming phenomena in animal groups have relied on real space dynamical 
models~\cite{Sumpter2010}.
Here, we address the issue of group formation using a network dynamical 
model ~\cite{Marsili2010,Ehrhardt2006}.
{neighbouring nodes  in the graph correspond to neighbouring individuals in space (Figure~\ref{fig:net_up_proc}). }

{Let us consider a group with $N$ individuals.  In our network model each 
individual is represented by a node (thus $N$ is the total number of nodes)} 
and 
each node $i$ has an internal dynamical variable $a_i$ that can take integer 
values ranging from $1$ to $q$. 
{Although the mathematical solution does not depend on the specific interpretation of the variable $a_i$, 
in the context of migrating groups, $a_i$ might be considered as the direction taken by a single individual 
to reach the destination site.}
{Links between nodes represent interactions among individuals by which they influence each other in their choice of the 
destination. While space is not explicitly resolved we assume that neighbouring nodes  in the graph correspond to 
neighbouring individuals in space (Figure~\ref{fig:net_up_proc}). Yet nearby individuals need not necessarily influence 
each other (see below).}

More precisely, a state of the system is defined by the adjacency matrix of the 
system $g_{ij}$  and by  the 
set of the internal dynamical states  variables $a_i$. {In our model, links are \emph{mutual} and, thus, the adjacency 
matrix is a symmetric matrix (i.e. for all $i, j$ we have that $g_{ij}=g_{ji}$) such that $g_{ij}=1$ if there is a link 
between the nodes $i$ and $j$,  $g_{ij}=0$ otherwise.} The evolution of  the system is governed by  
stochastic dynamics in which both the neighbourhood and the values of the internal dynamical state may vary, {according 
to stochastic Poissonian processes. These are discussed in the following (we refer the reader to the Supplementary 
Material for a detailed mathematical definition).}
 
\subsection{Network dynamics}
 
 The network evolves by creation and destruction of links, that mimic the spatial interaction between individuals. Link 
creation is quantified by the rate $\eta$ at which individuals form new links with other  individuals, 
This {rate } encodes both evolutionarily selected traits for 
pro-social behaviour and environmental factors, notably the 
 average distance between individuals. 
{In our model we assume that the interactions between individuals heading towards different directions decay 
much faster than interactions between close individuals heading towards the same direction. 
This is in agreement with real space dynamical model and is achieved in the network by assuming that link creation 
can occur only when individuals have the same internal state $a_i=a_j$. 
This is equivalent to saying that if $a_i\neq a_j$ the link between nodes decays immediately.} 

Finally, individuals {linked and} moving in the same direction can {also} move further apart from each other, which is 
formally encoded by assuming that links between nodes decay with a constant rate $\lambda$. {These two processes 
provide a {\em mean field} description of the real space dynamics.  Indeed link creation and decay depend on the 
geometry of the neighbourhood in spatially explicit models, which is averaged out within the mean field description. 
Mean field approximations such as this one work very well to capture the qualitative behaviour of complex systems. To 
set an analogy,  in a gas, one does not need to trace the trajectory of each molecule. It is enough to provide a 
"collision integral", that loosely speaking gives the probability that a particle moving in a certain direction will 
interact with a particle moving in a different direction. Here we are taking the same approach.}

\subsection{Internal state dynamics}

The {change} of the internal state is a Poissonian process that occurs with 
rate $\nu$ for each individual.
The choice of the destination $a_i$ is influenced by two factors:
 \emph{i)} pro-social behaviour, by which an individual keeps the same destination of their neighbours 
 and \emph{ ii)} memory, by which an isolated individual preferentially heads toward a destination $\alpha_i$ that is 
encoded in its memory. 

More precisely, when an individual updates its internal state, {\em i)} if it is linked to other individuals(s), it will 
{update} its internal state conforming to the state of the majority in its neighbourhood; 
i.e., the new state $a'_i$ is: 
\begin{equation}\label{eq:conform} 
a'_i=\underset{x}{\mathrm{argmax}}\big(\sum_{j}g_{ij}\delta_{x a_j}\big )
\end{equation}
{In this formula 
$\delta_{xy}$ is Kronecker delta 
function: i.e. if $x=y$, $\delta_{xy}=1$, otherwise $\delta_{xy}=0$.} {Again this rule is necessary within our mean 
field description of spatial interaction because if an individual were to chose a direction which is different from 
that of the (majority of the) group it is in, it would quickly move far apart and its links would decay.}

On the other hand, {\em ii)} if an individual is isolated (not linked), its choice of the internal state is influenced 
by its preference for a destination that is encoded in their memory. More precisely, we assume that each individual has 
a preferred value of the internal variable, let us call it $\alpha_i$. In the case of an internal state update event, an 
unlinked individual will pick up a state according to the following 
probability distribution:
\begin{equation}\label{eq:prob_update}
 \Prob(a_i=a)=\frac{\ee^{h_{\alpha_i}\delta_{a \alpha_i}}}{q-1+\ee^{h_{\alpha_i}}}
\end{equation}
where $h_{\alpha_i}$ is a parameter that measures the intensity of the 
preference. {This encodes, besides information processing and storage capabilities, also environmental factors related 
to the properties of a 
given feeding site, such as quantity of prey, water 
temperature, water quality etc.

The fraction of individuals with a preferred destination $\alpha$ is 
$n^\alpha$ but we also contemplate a fraction $n^0$ of ``uninformed'' individuals, that have no \emph{ a priori} 
preference for any memorized destination. We use the convention that uninformed individuals have $\alpha_i=0$  and 
$h_{\alpha_i}=0$. Therefore, uninformed individuals update their direction at random, which is 
described by Eq. (\ref{eq:prob_update}) with $h_{\alpha=0}=0$.}

Previous network approaches used a voter 
model update rule instead of a majority rule~\cite{Gross2008,Huepe2011,Couzin2011};  
this choice makes no qualitative difference in the stationary case, since our main results are based on a state space 
decomposition (see Supplementary Materials) that remains valid as long as the update rule promotes local uniformity. 
However, we expect detectable 
differences in the transient behaviour of these systems. Biologically, a 
majority rule captures the non-linearity of group behaviour.

{As in spatially explicit models, in our description individuals compromise about directional choices. The majority rule 
does not prohibit that an individual $i$ heading towards a given destination may change its route upon the encounter of 
another individual $j$. While this is not an elementary event described by the processes above, it can clearly occur as 
a composite event that entails the decay of all the links of $i$, an update of its choice and the formation of a link 
with $j$. The probability of this event is non-zero and it decreases with the number of individuals $i$ is interacting 
with, as one expects. }

{In some cases individuals in groups need
 to compromise between information gathering from the environment and social
 cohesion of the group~\cite{Ward2011,Krause1992} and thus some previous modelling approaches have assumed a trade-off 
between information capabilities and pro-social behaviour, in that informed individuals have a reduced tendency to 
follow their peers. The present modelling framework may be extended to encompass this situation also by making, for 
example $\eta$ take different values for informed and uninformed individuals. This generalization of the model leads to 
the same conclusions as those discussed below but it comes at the cost of more complex mathematics. 
In addition, there is no conclusive evidence, as far as we are aware of, that such trade-off really exist in 
populations of fishes (see e.g.~\cite{Miller2013}). We have however checked that adding these trade-offs to the 
model is inconsequential as far as the main results of the model discussed here is concerned, that is why we discuss 
these aspects in the supplementary materials. } 


\subsection{Invariant {distribution}}

Given the above transition rates, we can write down the master equation  (see the Supplementary Material)
and derive the invariant distribution which describes the stationary state. One key observation in this is that, since 
only links between nodes with the same internal state can be established, the process will converge to states where all 
links $(i,j)$ are 
between nodes with $a_i=a_j$. Any state with links $(i,j)$ connecting nodes with $a_i\neq a_j$ is transient, i.e. is 
not going to occur in the long run. This allows us to partition the states of the system into a transient class and a 
closed ergodic class. This ensures that the invariant distribution  is \emph{unique}.
It can be shown (see Supplementary Materials for the details) that the process 
satisfies detailed balance and the probability to observe state with a given 
network $\{g_{ij}\}$ and profile of choices $\{a_i\}$ in the stationary state, 
is given by:

\begin{equation}\label{eq:invariant_measure}
  \pi(\{g_{ij}\}, \{a_i\})=\frac{1}{\mathcal{Z}}\prod_{j<i}\ee^{\sum_i h_{\alpha_i} 
\delta_{a_i\alpha_i}}\bigg(\frac{2\eta\delta_{a_i a_j}}{\lambda(N-1)}\bigg)^{g_{ij}}.
 \end{equation}
where $\mathcal{Z}$ is the normalization constant. { In 
particular, when, for some $i$ and $j$  we have that $a_i\neq a_j$ and $g_{ij}=1$ , the invariant 
distribution is  zero. We are also 
assuming the convention that $0^0=1$.}

 Let $N_a^\alpha=\sum_{j}\delta_{a_j  a}\delta_{\alpha_j \alpha}$ be the number of 
 individuals that are in state  $a$ but would like to be in state $\alpha$ and let 
 $n_a^\alpha=\frac{N_a^\alpha}{N}$. 
 
If, in eq.~\eqref{eq:invariant_measure}, we call  
\begin{equation}\label{eq:sociality}z=2\frac{\eta}{\lambda} \end{equation} the non-dimensional parameter that accounts 
for the effective creation of links in the network, thus measuring  the \emph{sociality} of the group, 
{then} 
with standard mathematical manipulations (see Supplementary Materials) we can easily write the stationary state 
distribution  in terms of the densities $\bvec{n}=\{n_{a}^\alpha\}$ as follow:
  \begin{equation}\label{eq:invariant_measure_pop}
 p(\bvec{n})=\frac{1}{\mathcal{Z}}\ee^{-N \big[\fn{F}(\bvec{n};z,\bvec{h})+O(1/N)\big]}
\end{equation}
where
\begin{equation}\label{eq:free_energy}
\fn{F}(\bvec{n};z,\bvec{h})=\sum_a n_a^0 \log(n_a^0)+\sum_{a\alpha} n_a^{\alpha}\log(n_a^{\alpha})-\sum_{a \alpha } 
h_{\alpha} n_a^{\alpha}\delta_{\alpha a}-\frac{z}{2}\sum_a (n_a)^2
\end{equation}
and  $\mathcal{Z}$ is the normalization constant.
In the large population limit ($N\to \infty$) this distribution peaks exponentially in 
$N$ around the minima of $\fn{F}$.

The stationary  points of $\fn{F}(\bvec{n};z,\bvec{h})$ satisfies the following system of equations:
\begin{equation}\label{eq:FOC}
n_a^{\alpha}=\ee^{ h_{\alpha}\delta_{a\alpha}+z n_a} \frac{n^\alpha}{(\ee^{h_\alpha}-1)\ee^{z n_\alpha}+\sum_a \ee^{z 
n_a}}  
 \end{equation}
 where $n_a=\sum_i n_a^i$ is the total density of individuals whose internal state is $a$  (See Supplementary Materials 
for detailed calculation).

Therefore with a large number of individuals and in the stationary state of the system  
we are able to use Eq.~\eqref{eq:FOC} to analytically describe 
the fraction of individuals with {\em a priori}  preference $\alpha$ that end up heading towards 
destination $a$.

This set of non-linear equations has many solutions in principle. 
Those corresponding to stationary states can be fully characterized in terms of the average 
degree of the network $\langle k \rangle$ (i.e. the average number of neighbours of individuals) that is a 
proxy for the school density. It can be shown that one measure of the network degree 
is $\langle k \rangle=z\bigg(1-\frac{1}{q}\bigg)\sigma +\frac{z}{q}$ where the 
quantity:
\begin{equation}\label{eq:sigma}
   \sigma=\frac{q\sum_i (n_i)^2 -1}{q-1}.   
\end{equation}
is a direct measure of the school efficiency and it takes values between 
$\sigma=1$, when all individuals belong to a group that 
migrates towards the same destination; and $\sigma=0$ when individuals 
distribute equally between different destinations. 
Hence the solution with high coordination ($\sigma\simeq 1$) also corresponds to high 
network densities $\langle k\rangle \simeq z$.

Among all the solutions of Eq.~\eqref{eq:FOC}, we shall focus on those corresponding to the global minimum of 
$\fn{F}(\bvec{n};z,\bvec{h})$ that determine the behaviour of the system, since they correspond to the values around 
which the stationary distribution shall peak.

\section{Results}
We shall analyse two cases 1) the case of a population without informed individuals, $n^0=1$, and 
2) the case where a fraction $n^1=1-n^0$ of the individuals have a preferred migratory destination, 
whereas the rest is not informed.

\begin{figure}[!h]

\centering
\includegraphics{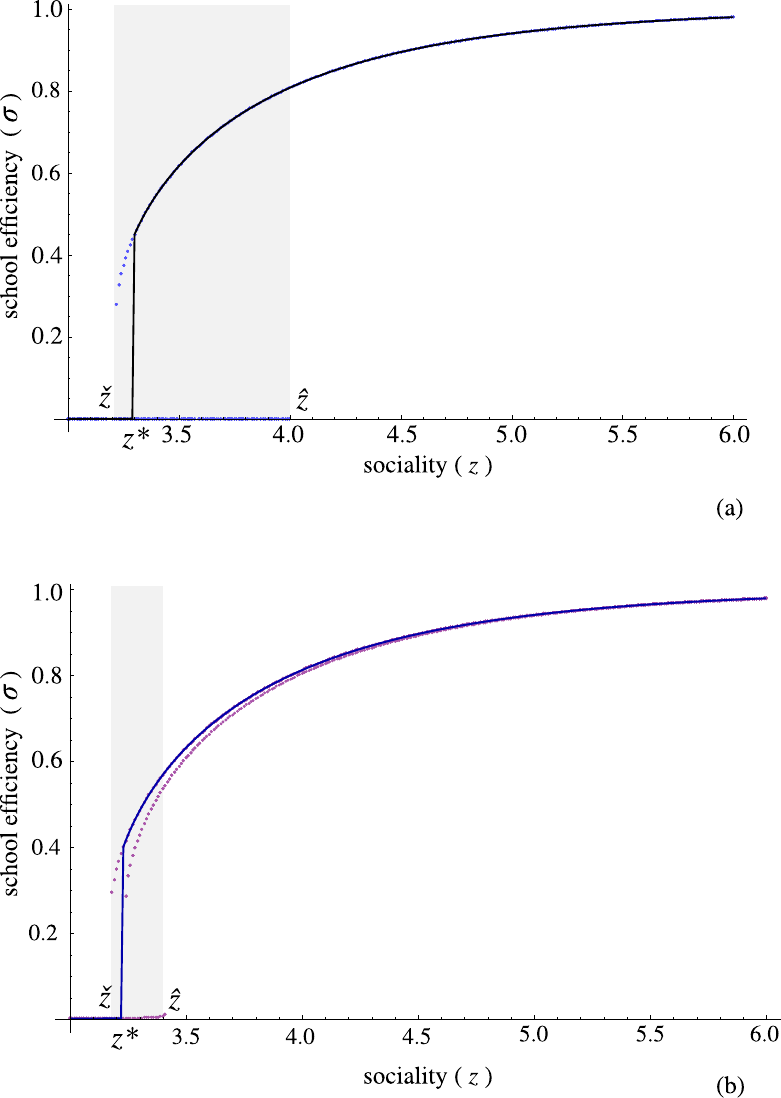}

\caption{Critical group dynamic: school efficiency, $\sigma$ as function of the social parameter $z$ in (a) non informed 
group $n^1=0$ and
 (b) informed group $n^1=0.05$, $h=0.5$. The dotted lines correspond to  all 
the stable solutions of~\eqref{eq:FOC}, the shadowed areas identify the 
coexistence region whereas the solid lines correspond to the equilibrium 
solution.}\label{fig:switch_n0n1}
\end{figure}

\subsection{Migration without information}

When no information is available in the group, the system reduces to an adaptive network model in which group 
coordination only depends on the rates at which links are created or destroyed~\cite{Marsili2010,Ehrhardt2006}.

Below a certain threshold $\check{z}$ only one local minimum exists which corresponds to a symmetric solution $\sigma=0$ 
(Figure~\ref{fig:switch_n0n1}~a);
there the network is sparse,$\langle k\rangle<1$, and the group does not migrate. At $\check{z}$, a new bundle of 
$q$ local minima appears at which $\sigma>0$. There the network is dense, $\langle k \rangle>1$, and a  fraction of the 
individuals comparable with $N$ (called in graph theory \emph{giant component}) is connected with one another  and 
coordinated on the same destination choice.

The analysis also produces the full probability distribution of different states that allows ranking the solutions in 
terms of their probability (see Supplementary Materials).
Between $\check{z}$ and $\hat{z}$ both solutions coexist and individuals can migrate in a coordinated manner or not.
Above $\hat{z}$ the only local minima are for $\sigma>0$ while the sparse solution $\sigma=0$ becomes unstable. There is 
an intermediate point $z^*$ below which the sparse solution is the most likely outcome whereas, above it, the high 
density solution will prevail.

\begin{figure}[!h]
 \centering
\includegraphics{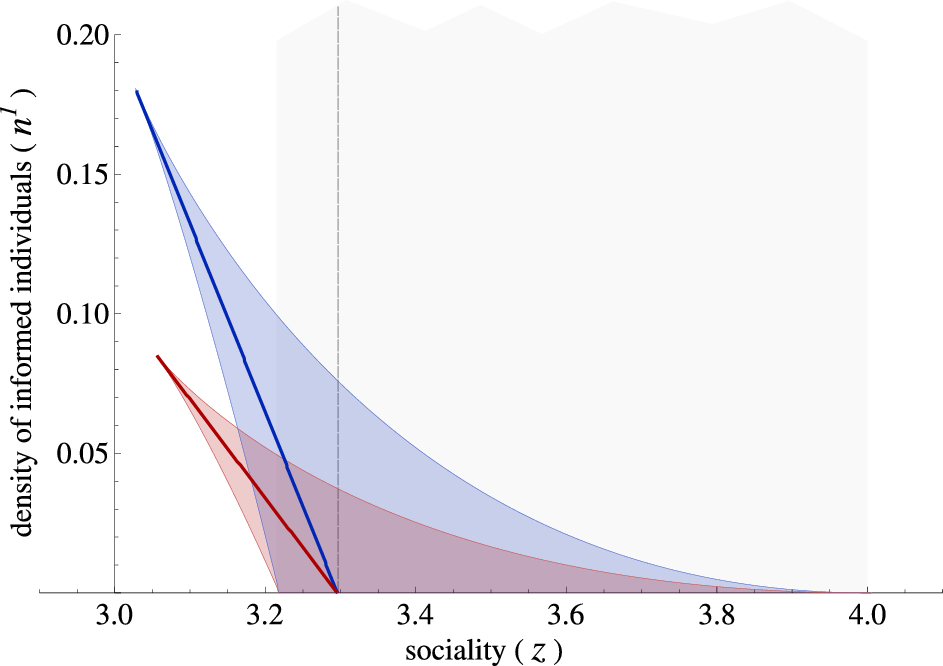}
\caption{Phase diagram of the system with $q=4$ possible directions.  The grey area corresponds to a preference 
parameter  $h=0$ (no preferences) and the dashed line is the critical line. The blue area corresponds to $h=0.5$ and the 
thick blue line represents the corresponding  critical line.  The red area corresponds to  $h=1$ and the thick red line 
represents the corresponding critical line as well.}\label{fig:phasediagram}
\end{figure}

\subsection{Informed migration}

In order to analyse the  role of information in the model, we study the simplest possible case, 
 with $q$ destinations, a density of informed individual $n^1=1-n^0$ and a preference $h$ about a single destination. 

The equation \eqref{eq:FOC} again can be solved numerically to obtain prediction on schooling behaviour. Information has 
two main effects on the system (Figure~\ref{fig:switch_n0n1}~b).
First, it breaks the symmetry between the $q$ high density solutions found in the $n_0=1$ case, by selecting the 
solution with the preferred destination $\alpha=1$ as the most likely.  The $q-1$ solutions  corresponding to migration 
toward other destinations remain stable, but are much less likely to be selected by the population. 

Secondly, the coexistence region between high and low density solutions $[\check{z},\hat{z}]$ 
is reduced in the case of informed migration (Figure~\ref{fig:switch_n0n1}). 
In fact this region becomes smaller as the number of informed individuals increases (Figure~\ref{fig:phasediagram}).

Eventually, there exists a critical value of $n^1$ at which the region collapses into a point. This change in the 
behaviour of the system is equivalent to a second order phase transition in physics. For values of $n^1$ greater than 
this critical point the system has a smooth transition between low and high density states, as $z$ increases, and a 
single solution is found.
Moreover the coexistence region and the critical value change with $h$. 
The thick line in Figure~\ref{fig:phasediagram} marks the point, in the coexistence region, where the two solutions are 
equally probable; on the right (left) of this line we expect to see the high (low) density solution.

\begin{figure}[!tbp]
 \centering
\includegraphics{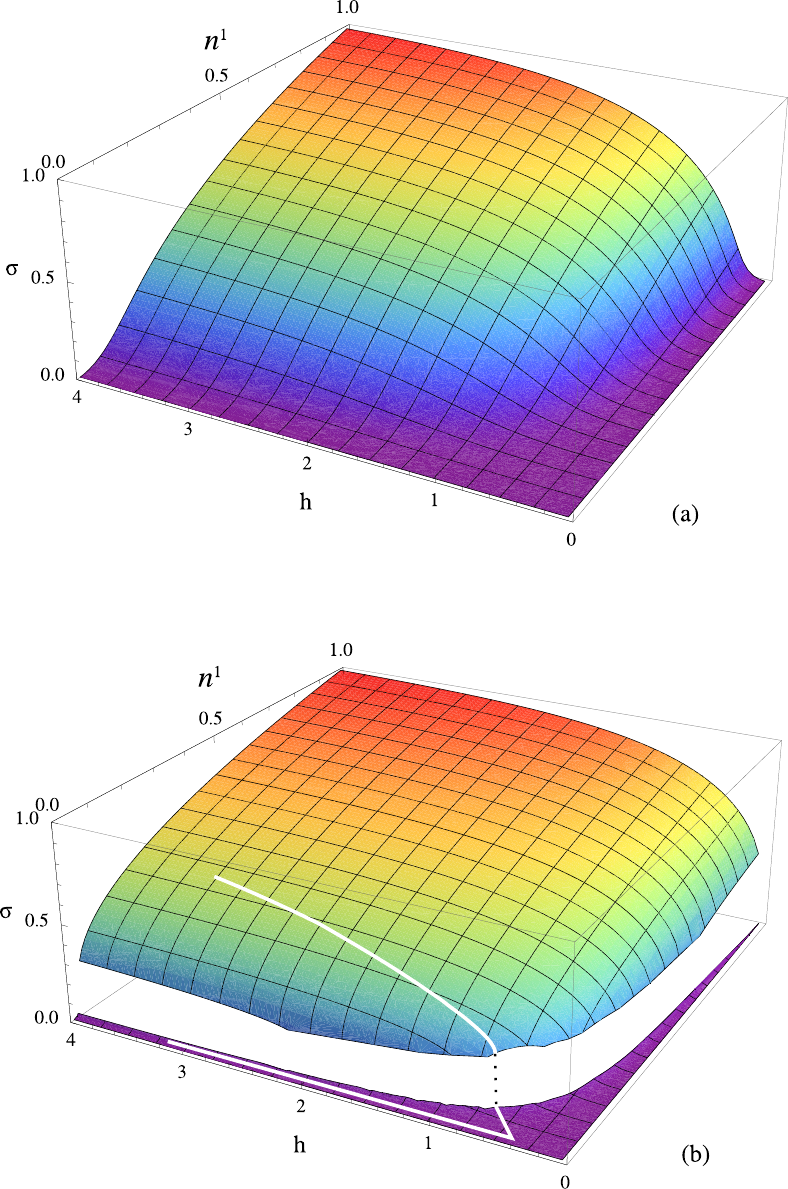}

\caption{School efficiency $\sigma$ as function of the fraction of informed individuals $n^1$ and strength of the 
preference $h$
when the social parameter $z$ is (a) in a non critical region $z=2.5$ and (b) in a critical region $z=3.1$ 
The white line in panel (b) is a schematic illustration of the hysteresis mechanism for a bluefin tuna population 
starting with high $n^1$ and $h$, then decreasing $n^1$ and $h$ (overfishing of both preys and predators) and 
subsequently increasing $h$ (increase of population of preys).  } \label{fig:mesh_znc_zc}
\end{figure}

The behaviour of the solution as the parameters $h$ and $n^1=1-n_0$ vary, at fixed $z$, is depicted in 
Figure~\ref{fig:mesh_znc_zc}. For low values of $z$ (Figure~\ref{fig:mesh_znc_zc}a) we observe a smooth crossover from 
low to high density solutions as $h$ and/or $n_1$ increase whereas when $z$ is larger the system exhibits a sharp 
transition between the two solutions (Figure \ref{fig:mesh_znc_zc}b). The presence of a sharp transition with 
coexistence in a broad range of parameters is a robust feature of this model.

For more complicated settings using competing groups with different preferred migratory destinations, it can be shown 
that, for large $z$ the population coordinates towards the migratory destination that provides the largest product 
$n^\alpha h_\alpha$ (see Supplementary Materials).
This quantity can be interpreted as the strength of the group's collective memory toward a given migration site, 
$\alpha$.

This provides us with a vivid picture of how we expect the collective behaviour of the population to change when the 
parameters $z$, $h$ and $n_0$ change. 
Adapting this picture to the observed behaviour of populations provides hints on the likely underlying causal effects. 
In brief, when $z$ is large, i.e. for individuals with a marked pro-social behaviour, we expect abrupt transitions when 
either the density $n^\alpha$ of individuals with a given preference, or the intensity $h_\alpha$ of that preference 
varies in such a way as to cross the boundaries in the phase diagram (Figure~\ref{fig:phasediagram}).

When both the density of informed individuals and the intensity of preference $h_\alpha$ decrease,
abrupt transition from efficient group formation to collapse of migration efficiency is visible. 
We note that this hysteresis cycle is consistent with observed stock collapses of migratory 
fish populations~\cite{Petitgas2010}. When the migratory population is described 
using a social parameter $z$ close to the critical point, then the interplay 
between the memory for a given destination, $h$,
and the fraction of the individuals informed, $n^1$ about this destination can produce an abrupt transition 
in the migration of the species.

In the case of a school migrating in direction $1$, a decrease of the value of $h$ and $n^1$ over years due, for 
example, to overfishing of both individuals and prey in the migration site, 
can force the system to cross the critical line reaching eventually low values of both $h$ and $n^1$. 
{When in this condition, an increase in the value of $h$ might occur due for example to better habitat conditions or 
food availability, 
for those few vagrant fish that might still be present in the area. However this increase alone cannot bring the system 
back to} 
the original state because the system may not cross again the critical line.
Thus the group may not migrate in direction $1$ even though previous 
{habitat} conditions are 
re-established.

\section{Discussion}
We show that abrupt changes in migratory patterns of animal groups can be caused by 
removal of knowledgeable individuals from the group or by decreasing preference of the individuals 
towards a particular migratory destination.{We demonstrate this with a robust analytical approach that allows to 
clearly 
identify the factors regulating group formation processes.} Our results are consistent with previous models suggesting 
that a small number 
of informed individual can lead to large group migrations~\cite{Huse2002,Couzin2005}. 
Additionally we demonstrate that diminishing individual preference for a given migration site 
can preclude group formation and break the migration process. 
\subsection{The migration game}
The migration process can be described as an emergent property of the population undertaking 
a  group formation game: when the spatial density of fish is locally low, each individual moves 
independently, and the system is in a sparse network configuration with a value of $z$ below the lower edge 
of the coexistence region.
In this state uninformed individuals cannot migrate whereas informed individuals can undertake 
a\emph{ solitary } migration towards their preferred destination. 
Owing to external stimuli (water temperature, local currents, topography,  etc. ) the density may increase 
and so does the value of $z$, driving the system toward the coexistence region. In this region even
 though the local density of fish is high, a sparse network configuration with fish moving 
 independently is still stable but an alternative and stable dense network configuration also appears. 
 When the system reaches the upper edge of this region, further increasing the 
density, the sparse network {state} becomes 
 unstable while the dense network {state} prevails and 
the school starts a migration toward the preferred destination. 
On the other hand an \emph{hysteretic} cycle is present in this system and when  the local density of fish decreases in 
the school, $z$,  decreases and the system is driven back to the coexistence region. 
A similar effect can be reproduced in the system by lowering the preference factor, $h$.
The schooling configuration remains stable until the system reaches the lower edge of the coexistence 
region:  at this point, fish stop schooling and the system switches back into the sparse configuration { (solitary 
fish)}. 

 The group formation game described above can be repeated each year naturally driving changes in 
the preference term $h$, hence in the memory of migratory fish. Likewise changes in this or in the 
other terms of the model may occur when the migratory population is affected by external stimuli
, e.g., overfishing, habitat degradation, demographic fluctuations. 
Because of the hysteretic cycle, such variations may then result in 
\emph{abrupt} changes in the migratory patterns.

\subsection{Conflicting preferences}
From the asymptotic analysis (Supplementary Materials) we demonstrate that, for large value of $z$, the 
group shall migrate toward the direction $\alpha$ for which  the product $n^\alpha h_\alpha$ is maximal, whereas in the 
limit of small $z$, the sparse configuration is the only stable one.
This suggest that our results might be extended to groups with conflicting preferences.
It is relevant to note that in our model \emph{all individuals} have a social component. For example in groups with 
conflicting preferences 
our model suggests that, for some  range of the parameters, an informed individual can follow the group 
and migrate toward a site different from {its} preferred destination. 
This approach makes our definition of leaders not only dependent on the amount of information stored but also on the 
social context in which they live. Therefore,  the interaction between personal information and social effects is 
explicitly resolved in our model 
and---we note---it has been suggested to operate in living groups~\cite{Miller2013,Herbert-Read2013}.

\subsection{Collective memory and breakdown of social traditions}
Breakdown of social  traditions, due to selected fishing on older informed individuals, has been 
hypothesized to have contributed to stock collapses in several large commercially important fish 
populations~\cite{Petitgas2010,Brown2003}. 

Our sketch of the migration game suggests that social dynamics may  lead to such collapses and that   the integrity of 
migration pathways and spatial distributions of migratory predators 
might be particularly vulnerable to perturbations such as fishing or habitat degradation.  
Fishing out informed individuals and their prey can exacerbate the loss of 
collective memory up to the point where a migratory pathway is suddenly 
interrupted.  
We can  assume that each year  young individuals  join the group: among them a fraction is able to gather information 
and remember a migratory route whereas the rest has a  purely social behaviour. 
The ``information-gathering-able'' individuals behave as uninformed individuals ($h=0$)  but \emph{ learn} a new 
migratory route during the first migration(s). If the group does not succeed in starting migration, or migrates toward a 
different location, the young ``information-gathering-able'' individuals will not learn the traditional migration route 
of the group and the social traditions of the group will not be transmitted to the new generations.
The loss of collective memory in the group will then force the system to cross the critical line and the migration 
toward the destination site will stop.

An example of a prey-predator collapse and subsequent abrupt disappearance of migratory route is provided by Atlantic 
bluefin tuna (\emph{Thunnus thynnus} \textsc{Linnaeus},  1758) 
and its main prey, {herring (\emph{Clupea harengus} \textsc{Linnaeus}, 1758)  in the Norwegian and the North Seas.} 
During the 1950s-1970s both species were heavily exploited in {these regions}
resulting in the disappearances of both species~\cite{ICES2011,ICES2012, ICCAT2012}.  
Since then, the herring populations in both regions have recovered to moderate-high levels 
~\cite{ICES2011,ICES2012}, but bluefin tuna {have been extremely rare during the 1980s-2000s and apparently had 
not 
migrated to 
these areas in large numbers since the disappearance several decades ago ~\cite{ICCAT2012}. }
These hysteretic dynamics are consistent with a fishing-induced removal of predators  having 
preference for migration to these regions and a fishing induced decline in habitat quality which 
then leads to the collapse of group formation and a sudden change in migratory 
path (cf. Figure~\ref{fig:mesh_znc_zc}).

\section{Conclusions}
We have presented a model that offers and elucidates 
a plausible mechanism for migration dynamics.  
By extending and generalizing previous approaches, our model shows that 
group formation dynamics have a critical dependence on both sociality, number 
of informed individuals and strength of the preference in informed individuals. 
For example, partial removal of knowledgeable individuals  may be sufficient to interrupt the transmission of social 
traditions in groups of animals.  
Such critical dependence is consistent with abrupt transitions that are commonly 
observed in migration patterns of social animals such as Atlantic bluefin tuna 
{as well as other fish populations~\cite{Petitgas2010}.

Our findings offer deep insight into migration dynamics and suggest interesting directions both for data analysis 
(e.g. new interpretations of spatial temporal dynamics of migratory populations)
and for further theoretical development (e.g. accounting for conflicting preferences, continuous directions, different 
segregation policies, topological interaction).}
{Contrary to previous Agent Based approaches \cite{Katz2011,Vicsek2012,Sumpter2010,Chate2008,Guttal2010}, our model has 
the advantage of being analytically soluble, and thus it provides a powerful theoretical bench test for hypotheses on 
collective animal behaviour.}

\section*{Acknowledgements}
We thank the Danish National Research Foundation (Dansk Grundforskningsfond) for 
support to the Centre for Macroecology, Evolution and Climate, University of 
Copenhagen. { Part of the research leading to these results has received the support by the EU-FP7 project 
EURO-BASIN (grant agreement n\textsuperscript{o} 264933). } PM was also supported by the 
project  "North Atlantic - Arctic coupling 
in a changing climate(NAACOS)" funded by the Danish Council for Strategic 
Research.

\newpage

\appendix

\section{Supplementary materials}

In these supplementary materials we provide the details of the technical calculation that are presented in the main 
paper and provide additional {comments and details.}
 
In the first section we detail the definition of the model. In the following section we comment on the interpretation 
of the parameters in terms of fish migratory behaviour. 
The third section contains a detailed derivation of the stationary state distribution (the invariant measure), 
equations~(6) of the paper. From this we derive the population distribution Eq.~(8).

This is then analysed in the limit of large populations (the thermodynamic limit) leading to the expression of the free 
energy~(9)
of main paper. Finally we provide details on the calculation of the equilibrium solution~(10)  
of main paper, and of other results cited in the main paper.

\subsection{Mathematical definition of the model}\label{sec:model}

Let $N$ be the number of individuals in the group and $q$ the number of 
possible {values for that the internal variable can take (i.e. the 
possible directions a fish may take)}.

For any finite $N$ and $q$, {a 
\emph{state} of the system is defined by the network of interaction between 
individuals and by the values of the internal variables of  the individuals. 
The processes described in the main text define a stochastic dynamics on this 
state space.  For example, when a link is created between two individuals $i$ 
and $j$ the system will make a \emph{transition} between a state in which the 
network of interaction has no link between $i$ and $j$ and a state in which the 
link between $i$ and $j$ is added. We use the letter $\omega$ to generically 
refer to a state, i.e. } $\omega=(G,\bvec{a})$ where  $G$ is the $N \times N$ 
adjacency matrix of the system (i.e. $(G)_{ij}=g_{ij}\in\{0,1\}$), and 
$\mathbf{a}=(a_i)$ is a vector whose $i$-th component
is the values of the internal variable of $i$-th node. {We also use   
$\sset{S}$ to refer of all the possible states in which the system 
can be (i.e. the state space).} {
$\hat\omega(t)$ represents the state of the system at time $t$} which
 {shall} be equal to one of the states described 
above. {Mathematically, our system is \emph{a Continuous Time Markov Chain} 
and therefore, its evolution  over time} is described by a Master Equation for the probabilities:
\begin{equation}\label{eq:master_equation}
\partial_t P(\omega,t)=\sum_{\omega'}P(\omega',t)\rho(\omega'\to \omega)-P(\omega,t)\sum_{\omega'}\rho(\omega\to 
\omega')
\end{equation}
where
\begin{equation}\label{eq:p_omega}
P(\omega,t)=\Prob(\hat\omega(t)=\omega);
\end{equation} and the $\rho$ are the transition rates which correspond to the three dynamical processes described in 
Methods Section of the main paper. For clarity sake we describe them again here.
\begin{description}
\item[link creation]
With a rate $\eta$ each node $i$ can establish a link with another $j$ node picked up randomly among the others.
The link is established only if
$a_i=a_j$

This process can connect only two states $\omega$ and $\omega'$ such that 
$g_{hk}=g'_{hk}\  \forall\ \!\  (h,k)\neq(i,j)$, $g_{ij}=0$ and $g'_{ij}=1$. The 
transition rate is clearly
\begin{equation}\label{eq:linkcreation}
\rho(\omega\to\omega')=\frac{2\eta}{N-1}\delta_{a'_i a'_j} 
\end{equation}

\item[link destruction]
Each link ha a destruction rate $\lambda$. 

This process can connect only two states $\omega$ and $\omega'$ such that 
$g_{hk}=g'_{hk}\  \forall\ \!\  (h,k)\neq(i,j)$, $g_{ij}=1$ and $g'_{ij}=0$. The 
transition rate is clearly
\begin{equation}\label{eq:linkdestruction}
\rho(\omega\to\omega')=\lambda 
\end{equation}

\item[preference update]

Each node can update its internal state at a rate $\nu$.

If the node is linked when an internal state update event occurs, it will 
conform to its neighbourhood. {When an internal state update event occurs 
for a linked individual $i$, thus, the new internal state of that individual 
$a'_i$ is chosen using a \emph{majority rule}, that is :
\begin{equation}\label{eq:conform}
 a'_i=\underset{x}{\mathrm{argmax}}\big(\sum_{j}g_{ij}\delta_{x a_j}\big )
\end{equation}}

This assumption is coherent with what is usually done in modelling group motion 
that is assuming  that an individual tend to
"follow" its neighbours. 

Instead, when a node is not linked, it undergoes a random transition to a 
state $a'_i$;  the probability 
 of picking up one direction over another encodes the \emph{a priori} 
information the individual has.
 Each individual has a direction preference $\alpha_i$. If $\alpha_i=0$ then 
each 
 direction has the same probability($\frac{1}{q}$) of being chosen (no \emph{a 
priori} information); if $\alpha_i\in \{1,...,q\}$ 
 than the $i$-th individual has a higher probability of picking the direction 
$\alpha_i$ over the others. {In mathematical term, this can be written as:
 \begin{equation}\label{eq:unlinked}
   \Prob(a'_i=a)=\frac{\ee^{h_{\alpha_i}\delta_{a 
\alpha_i}}}{q-1+\ee^{h_{\alpha_i}}},~~~g_{ij}=0~\forall j
 \end{equation}
where $h_{\alpha_i}>0$ measure the strength of the preference of node $i$ for the 
direction $\alpha_i$.
 }
We assume that the strength  of the preferences $h_\alpha$ does not vary among 
the individuals that share the same direction preference but may be different
for individuals preferring different directions.

\end{description}

{
All other transition besides the three classes discussed above, have zero rates.

The model can be generalized in different ways. In particular, the creation rate $\eta^\alpha$ can be taken to depend 
on whether individuals are informed ($\alpha>0$) or not ($\alpha=0$). 
In general, if the propensity for forming social links among fishes with preferred direction $\alpha$ is $\eta^\alpha$, 
then this more general model entails substituting $2\eta$ with $\eta^{\alpha_i}+\eta^{\alpha_j}$ in 
\eqref{eq:linkcreation}. This reflects the fact that the creation of the link between $i$ and $j$ may be initiated by 
either $i$ or $j$ an hence their rates add. 

For the sake of simplicity, we shall focus our discussion to the case $\eta^\alpha=\eta$ for all $\alpha=1,\ldots,q$. 
Indeed, the derivation proceeds along exactly the same lines and the gist of 
the main results is the same. We shall deal with the general case in Section \ref{sec:tradeoff}.

\subsection{Interpretation of the parameters of the model}

Our theoretical framework is a stylised representation of  the migratory behaviour of fish populations.  
The three parameters of the model have a clear interpretation in terms of biological traits of the individuals and of 
the physical conditions of the environment in which they interact.

The value of $\eta$ must have a dependence on the density of 
individuals; most of the real space models assume that individuals interact only 
with ``close''  individuals (e.g. closer than a certain radius) which is a 
reasonable description of the natural behaviour of schooling fishes; thus if the 
local density of individuals is too low the probability of being close enough to 
interact with one another is small and thus the link creation rate must also be 
small; on the contrary the higher the density, the higher the number of 
``close''  individuals and the higher the creation rate must be. At the same time,
$\eta$ quantifies also the pro-social behavior of individuals which is encoded in 
the genetic make-up of the species. A social fish is expected to be able 
to interact with other individuals of its species more effectively  than a fish 
of a non social species: by the rules of our model, this means that at a given 
local density of individuals the social species will 
have a significantly higher link creation rate $\eta$ and lower link destruction rate $\lambda$
and thus a higher value of their ratio $z=\eta/\lambda$. A variation of local density may induce even 
significant variation of $\eta$ and therefore of $z$. The range of these 
variations,however,  may be seen as a \emph{genetically determined quantity}.

The interpretation of the other parameters is easier, the value of $n^\alpha$ is the 
relative proportion of individuals in the population with preference for destination $\alpha$; the value of $h_\alpha$ 
implicitly measures the strength of the preference, which in turn may encode the property of a 
given feeding site, such as quantity of prey, water 
temperature, water quality etc.

\subsection{Master Equation and Invariant Measure}\label{sec:stationary_state}

The derivation of the stationary state distribution, which is the solution of Eq. (\ref{eq:master_equation}) with 
$\partial_t P(\omega,t)=0$, relies on the following observations:

\begin{enumerate}[1.]
 \item the states of the system can be classified into 
two sets: one $\sset{A}$ that contains all the states  in which the network \emph{does not contain links between 
different nodes -- i.e. nodes that have different value of their internal variable --} and $\sset{T}$ containing all 
the states in 
$\sset{S}$ that are not in $\sset{A}$. In particular the states for which the 
network has no links are in $\sset{A}$;
\item transition from any state $\omega'\in \sset{A}$ to any state in $\omega\in \sset{T}$ are impossible since the 
rates vanish
\begin{equation}\label{eq:transient}
 \rho(\omega\to\omega')=0,\qquad \forall\omega'\in \sset{A},~~~\hbox{and}~~~\forall\omega\in \sset{T}.
\end{equation}
In words, link between nodes $i$ and $j$ with different internal variables $a_i\neq a_j$ cannot be generated in the 
course of the dynamics. 
\item If any such link exists at a given time $t$, then {\em i)} it must be present in the initial conditions and {\em 
ii)} it has a finite life time, because any link decays at a rate $\lambda$. 
\item As a consequence, the dynamics sooner or later reaches a state $\omega\in\sset{A}$ where no link between 
different nodes exists, and from that time onwards states $\omega'\in\sset{T}$ where at least one link between different 
nodes exists, will never be reached. In formal terms, this means that states $\omega\in \sset{T}$ are transient and 
therefore they occur with zero  probability in the stationary state
\begin{equation}\label{eq:transient}
 \lim_{t\to\infty}P(\omega,t)=0,\qquad \forall\omega\in \sset{T}.
\end{equation}
\item it is possible to reach any state in $\sset{A}$ starting from any state in 
$\sset{A}$. More precisely, for any two states $\omega,\omega'\in\sset{A}$ it is possible to find a sequence of  
transitions between intermediate states connecting $\omega$ to $\omega'$, with each transition having a positive
probability. One such ``path'' of transitions, for example'' is the one where first all the links in the initial state 
$\omega$ 
decay, then the internal variables of each node is updated from the one prevailing in state $\omega$ to the one in 
$\omega'$, and finally all the links in state $\omega'$ are sequentially added. Each of the intermediate states 
$\omega''$ along this path is also in $\sset{A}$ and each transition between consecutive states on the path occurs with 
a strictly positive rate. Since the number of states on the path is finite, this means that the probability 
$P\{\hat\omega(t)=\omega|\hat\omega(t_0)=\omega'\}$ to find the system in $\omega$ at time $t$ given that it was in 
state $\omega'$ at an earlier time $t_0<t$ is strictly positive.
\item Since the process can reach any state in $\sset{A}$ from any other state in $\sset{A}$ this proves that the 
dynamics is \emph{ergodic} when restricted to $\sset{A}$. This implies that the stationary state exists and is unique 
(see e.g. \cite{Norris1998}). 
\item it is easy to verify, by direct substitution, that the probability distribution
\begin{equation}\label{eq:invariantmeasure}\boxed{
\pi(\omega)=\frac{1}{\sset{Z}}
\left\{\begin{array}{cc}
\ee^{\sum_i h_{\alpha_i} 
\delta_{a_i\alpha_i}}\prod_{j<i}\left(\frac{2\eta
}{\lambda(N-1)}\right)^{g_{ij}} & \omega\in\sset{A} \\0 & \omega\in\sset{T}\end{array}\right.
}\end{equation}
where $\sset{Z}$ is the normalization constant ensuring $\sum_\omega 
\pi(\omega)=1$, satisfies the detailed 
balance condition
\begin{equation}\label{eq:detbal}
 \pi(\omega)\rho(\omega \to \omega')=\pi(\omega')\rho(\omega' \to \omega).
\end{equation}
Indeed, for each $\omega,\omega'$ for which $\rho(\omega \to \omega')=0$ we have either that 
$\rho(\omega' \to \omega)=0$ also or that $\rho(\omega' \to \omega)>0$ but $\omega'\in\sset{T}$, and therefore 
Eq. (\ref{eq:detbal}) holds because $\pi(\omega')=0$. 
When both $\rho(\omega \to \omega')>0$ and $\rho(\omega' \to \omega)>0$, then either $\omega,\omega'\in\sset{T}$ 
and then Eq. (\ref{eq:detbal}) holds because $\pi(\omega)=\pi(\omega')=0$, or $\omega,\omega'\in\sset{A}$. In the 
latter case $\omega$ differs from $\omega'$ either for the presence of one link or for the value of the internal 
variable of a single isolated node. In both cases, 
one can check that Eq. (\ref{eq:detbal}) holds. This means that if $P(\omega,t)=\pi(\omega)$ then $\partial_t 
P(\omega,t)=0$, i.e. $\pi(\omega)$ is a stationary distribution of the process.
\item Since the stationary state is unique we conclude that 
\begin{equation}
\pi(\omega)=\lim_{t\to \infty}P(\omega,t)
\end{equation}
is the invariant {distribution}.
\end{enumerate}

The fact that a process satisfies detailed balance is related to the existence of a potential function 
$\sset{H}(\omega)=-\log\pi(\omega)$ such that each transition can be interpreted as either ``climbing'' or 
``descending'' the landscape of $\sset{H}$. 
In order to gain intuition on why the process satisfies detailed balance one can argue that such a function $\sset{H}$ 
exists for this process. Indeed, notice that the process links states which differ by one link (with the profile 
$\bvec{a}$ of nodes' variables constant) or states with the same graph $G$, which differ only by the attribute $a_i$ of 
a single isolated node. In the fist case, the transition rate is the same, so the process, at fixed $\bvec{a}$, can be 
described as 
``climbing'' or ``descending'' a step of a function $\sset{H}$ that depends only on the number of links added.
In the second case, the dynamics of $\bvec{a}$ at fixed $G$ involve rates that depend only on whether a node is updated 
so that his attribute $a_i$ equals the preferred value $\alpha_i$ or not. This can be captured by a potential 
$\sset{H}$ that takes two different values depending on whether $a_i=\alpha_i$ or not.}

{
We note that $\pi(\omega)$ is the invariant distribution under a broader set of choices of the internal state update 
rule. 
Indeed Eq. (\ref{eq:conform}) can be replaced by any rule by which the choice of the internal state $a_i'$ is limited 
to the values $a_j$ of the neighbours of node $i$. For example, if Eq. (\ref{eq:conform}) is replaced by a rule where 
node $i$ takes the value $a_j$ of a randomly chosen neighbour $j$, as in the voter model, the process converges to the 
same invariant distribution $\pi$. This is because once the process reaches a state $\omega\in\sset{A}$, the link 
creation/destruction policy ensures that local uniformity that characterizes states in $\sset{A}$ will be preserved. 

We may expect, however, that the choice of the internal state update rule may influence the transient behaviour of the 
system (e.g. average time to reach the stationary state, structure of the metastable states).The analysis of those 
behaviour, albeit very interesting, is beyond the scope of this work.
}

\subsubsection{The population distribution}
In order to take the thermodynamic limit we need to obtain an expression in 
which the number of nodes $N$ is explicit. That is we need to infer from the invariant 
measure~\eqref{eq:invariantmeasure} an expression for the probability 
distribution function of the populations involved.

Let's consider a partition of the entire population (i.e. the number of nodes) 
$N$ into  classes. Let $N_a^{\alpha }=|\{i|(a_i=a) \wedge (\alpha_i=\alpha)\}|$ be the number of nodes with preferred 
state $\alpha$ that are in the state $a$.
Moreover we define a class of uninformed individuals and define $N_a^{0}=|\{i|(a_i=a)\wedge(\alpha_i=0)\}|$ as the 
number of nodes with state $a$ that have no preference (i.e. 
$\alpha=0$). We denote by  $\mathbf{N}=\{N_a^\alpha,~a=1,\ldots,q,~\alpha=0,1,\ldots,q\}$ the profile of population 
occupation states. 
Obviously
\begin{equation}\label{eq:normalization2}
 \sum_{a,\alpha }N_a^{\alpha }+ \sum_a N_a^0=N
\end{equation}

We define 
$\hat{\Omega}(\mathbf{N})=\{ \omega \in (G,\bvec{a}) : (|\{j : (a_j=a) \wedge 
(\alpha_j = \alpha) \}|=N_a^{\alpha })\wedge(\{j : (a_j=a)  \wedge 
(\alpha_n=0)\}|=N_a^{0})\}$ as the subset of states $\omega$ with profile $\bvec{N}$. 
The probability to observe $\bvec{N}$ is clearly  given by
\begin{equation}
\label{ }
p(\bvec{N})=\pi(\hat{\Omega}(\bvec{N}))=\sum_{\omega\in\hat\Omega(\bvec{N})}\pi(\omega).
\end{equation}

For sake of simplification we shall use  the following notation: 
$N_a=\sum_{\alpha }N_a^{\alpha  }+ N_a^0$ which denotes the number of nodes in 
actual state $a$,  $N^{\alpha}=\sum_{a  }N_a^{\alpha  }$ which denotes the 
number of nodes with preferred state $\alpha$, $N^0=\sum_{a }N_a^{0}$ which 
denote the number of uninformed nodes. Clearly $N^{\alpha}$, 
$N^{0}$ are known and fixed.

For each configuration of the internal variables, the equilibrium dynamics 
allows $\sum_{a} 2^{\frac{N_{a}(N_{a}-1)}{2}}$ different network configurations 
with non-zero probability. The weight of all network structures with  
a given number $m_a$ of links in a given preference class  is the same 
(since only networks 
with links between coordinated nodes carry non zero contribution). Calculating 
their contribution to $p(\bvec{N})$ reduces to counting  how many different 
network structure are there with a given number of links  $m_a$ in each component $a$. The answer is 
trivially $\binom{{\binom{N_a}{2}}}{m_a}$ The network contribution to 
$p(\bvec{N})$ is then
\begin{equation}
 \sum_{m_1,\ldots, m_q} 
\prod_{a=1}^q\binom{\binom{N_{a}}{2}}{m_a} \left[\frac{2 
\eta}{\lambda(N-1)}\right]^{m_{a}}=\prod_{a}\left[1+\frac{2 
\eta}{\lambda(N-1)}\right]^{\binom{N_{a}}{2}}
\end{equation}

Concerning the statistical weight coming from individual preferences, 
 we notice that:
 \[
\sum_i h_{\alpha_i}\delta_{a_i \alpha_i} =\sum_\alpha h_\alpha\sum_i\delta_{\alpha_i,\alpha}\delta_{a_i \alpha}
= \sum_{\alpha}h_{\alpha}  N_\alpha^{\alpha}
\]
Thus this results in a statistical weight given by $\exp(\sum_{a} h_{a}  N_a^{a})$. 

In order to compute $p(N)$ we notice that there are exactly 
$\frac{N!}{\prod_a N_a^0!\prod_{a\alpha  }N_a^{\alpha}!}$ configuration with the 
same weight and thus we can write
\begin{equation}
 p(\bvec{N})=\frac{1}{\sset{Z}}\frac{N!}{\prod_a N_a^0!\prod_{a\alpha  
}N_a^{\alpha}!}\ee^{ \sum_\alpha h_{\alpha} 
N_\alpha^{\alpha}}\prod_{a}\left[1+\frac{2 
\eta}{\lambda(N-1)}\right]^{\frac{N_{a}}{2}(N_a-1)}
\end{equation}
Where $\sset{Z}$ is the normalization constant.

The measure is defined over the multisymplex defined by:
\begin{equation}
0\leq N_a^\alpha\leq N^\alpha
\end{equation}
and 
\begin{equation}
\sum N_a^\alpha=N^\alpha
\end{equation}

\subsection{Thermodynamic limit and equilibrium solution}\label{sm:tl}

Once we have a form for the $p(\bvec N)$ we look for the asymptotic behavior for large $N$.

Let us denote the densities with $n_a^{\alpha}=\frac{N_a^{\alpha}}{N}$ and 
consequently $n_a=\sum_{\alpha} n_a^{\alpha}$ and $n^{\alpha}=\sum_a 
n_a^{\alpha}$ for $\alpha=0,1,\ldots,q$ and $a=1,\ldots,q$.

Next, we expand $\log p(\bvec{N})$ in a large N limit, using Stirling 
approximation($\log(N!)\simeq N\log(N)-N$, $\log(1+x)\simeq x$ for small $x$ and 
normalization :
\begin{equation}\label{eq:frqnormalization}
 \sum_{a}n_a^{\alpha}=n^\alpha.
\end{equation}
We then obtain :
\begin{equation}
 \begin{split}
\log(p(\bvec{N}))&=N\log(N)-N- \sum_{a\alpha}(N n_a^{\alpha}\log(N)+ N 
n_a^{\alpha}\log(n_a^{\alpha})-Nn_a^{\alpha})\\
+&N\sum_{a\alpha} h_\alpha 
\delta_{a\alpha} n_a^{\alpha}+ N\frac{\eta}{\lambda}\sum_a \bigg(\frac{N}{N-1} 
(n_a)^2+\frac{N}{N-1}\frac{1}{N} n_a\bigg)-\log{\sset{Z}}\\
=-&N\left[ 
\sum_{a\alpha}n_a^{\alpha}\log(n_a^{\alpha})-\sum_{a\alpha} h_\alpha  
n_a^{\alpha}\delta_{a\alpha} -\frac{\eta}{\lambda}\sum_a(n_a)^2+O(1/N)\right]-\log{\sset{Z}}
 \end{split}
\end{equation}

This expression is reminiscent of Gibbs distribution in statistical physics 
\begin{equation}
p(N)=\frac{1}{\sset{Z}}\ee^{-N 
\left[\fn{F}(\bvec{n};\frac{\eta}{\lambda},h)+O(1/N)\right]}. 
\end{equation}
where the \emph{free energy} is given by
\begin{equation}\label{eq:FreeEnergy2}
\fn{F}(\bvec{n};\frac{\eta}{\lambda},\bvec{h})=\sum_a n_a^0 \log n_a^0+\sum_{a 
\alpha} n_a^{\alpha}\log(n_a^{\alpha})-\sum_{a \alpha } h_{\alpha} n_a^{\alpha}	
\delta_{\alpha a}-\frac{\eta}{\lambda} \sum_a (n_a)^2 
\end{equation}

In the limit of very large $N$ the invariant measure concentrates on the global minimum $\bvec{n}^*$ of the 
function $\fn{F}(\bvec{n};\frac{\eta}{\lambda},h)$. Any state 
$\bvec{n}$ with a value of $\fn{F}$ that is larger by $\delta\fn{F}$ than the minimum, will have a probability 
$p(\bvec{n})\sim e^{-N\delta \fn{F}}$ which is exponentially small compared to $\bvec{n}^*$, i.e. will virtually 
never occur for large $N$. 

\subsubsection{Minimization of Free Energy}
We then have to minimize the free energy $\fn{F}(\bvec{n};\frac{\eta}{\lambda},\bvec{h})$ of the system, over the 
variables $\bvec{n}$ subject to the constraints:
 \begin{equation}\label{eq:constraint-2-1}
n^{\alpha}=\sum_{a}n_a^{\alpha} 
\end{equation}
and
\begin{equation}\label{eq:constraint-2-2}
n^0=\sum_a n_a^0 
\end{equation}

We introduce then Lagrange multiplier $\beta_{\alpha}-1$ for the first 
constraints and $\beta_0-1$ for the second one and impose first order conditions (FOC):
\begin{equation}\label{eq:FOC-2-1}
\nabla\left\{\fn{F}(\bvec{n};\frac{\eta}{\lambda},\bvec{h})-\sum_\alpha 
(\beta_{\alpha}-1)\left[-n^{\alpha}+\sum_{a} n_a^{\alpha}\right] -(\beta_0-1) 
(n^0-\sum_a n_a^0)\right\}=0 
\end{equation}
obtaining
\begin{equation}\label{eq:FOC-2-2}
\log(n_a^0)-\frac{2\eta}{\lambda}n_a-\beta_0=0
\end{equation}
and
\begin{equation}\label{eq:FOC-2-2}
\log(n_a^{\alpha})- 
h_{\alpha}\delta_{a\alpha}-\frac{2\eta}{\lambda}n_a-\beta_{\alpha}=0
\end{equation}
 and thus
\begin{equation}\label{eq:FOC-2-3}
\begin{cases}
n_a^0=\ee^{\frac{2\eta}{\lambda}n_a} \ee^{\beta_0} &  \\
n_a^{\alpha}=\ee^{ h_{\alpha}\delta_{a\alpha}+\frac{2\eta}{\lambda}n_a} 
\ee^{\beta_{\alpha}} & 
\end{cases}
\end{equation}

If we define :
\begin{equation}
Q=\sum_a \ee^{z n_a}
\end{equation}
we can write
\begin{equation}\label{eq:norm0}
 n^0=\ee^{\beta_0} Q
\end{equation}
and
\begin{equation}\label{eq:norm1}
n^{\alpha}= 
\ee^{\beta_\alpha}\ee^{\frac{2\eta}{\lambda}n_{\alpha}}(\ee^{h_{\alpha}}-1)+\ee^
{\beta_\alpha}Q
\end{equation}

Equation~\eqref{eq:norm0} allows us to eliminate one Lagrangian multiplier:
\begin{equation}\label{eq:zeta0}
 \ee^{\beta_0}=\frac{n^0}{Q}
\end{equation}

Equation~\eqref{eq:norm1} instead allows us to eliminate the quantity 
\begin{equation}\label{eq:betazeta} 
\ee^{\beta_{\alpha}}=\frac{n^{\alpha}}{(\ee^{h_\alpha}-1)\ee^{\frac{2\eta}{
\lambda}n_{\alpha}}+Q}
\end{equation}
We then have also the normalization constraint:
\begin{equation}
\sum_i n^i =1-n^0.
\end{equation}

We can then write the FOC for our system in the following way

\begin{equation}\label{eq:foc_def}
\boxed{
n_a=\ee^{\frac{2 
\eta}{\lambda}n_a}\left[\frac{n^0}{Q}+W+\frac{(\ee^h-1)n^a}{(\ee^h-1)\ee^{\frac{
2 \eta}{\lambda}n_a}+Q}\right]
}
\end{equation}

where 
\begin{equation}
Q=\sum_{i=1}^{q}\ee^{\frac{2 \eta}{\lambda}n_i} 
\end{equation}
and
\begin{equation}
W=\sum_{i=1}^{q}\frac{n^i}{(\ee^{h_i}-1)\ee^{\frac{2 \eta}{\lambda}n_i} +Q} 
\end{equation}
These equations can be solved numerically to any preassigned degree of precision.

\subsection{The $n^1=1-n^0$ case }\label{sm:1f}
When we consider  systems where only one direction is preferred, some 
simplification
can be made.

To ease the notation we shall write $z=\frac{2 \eta}{\lambda}$,\,$x=n_1$ and 
$y_i-1=n_i$ for $i\in\{2,..,q\}$.

In this case equation~\eqref{eq:foc_def} take the simplified form:
\begin{equation}\label{eq:focforcing}
\begin{cases}
x\ee^{-x}=z\frac{n^0}{Q}+z\ee^h\frac{1-n^0}{(\ee^h-1)\ee^x+Q} \\ 
y_i\ee^{-y_i}=zP=z\frac{n^0}{Q}+z\frac{1-n^0}{(\ee^h-1)\ee^x+Q} & \text{q-1 
times}
\end{cases}
\end{equation}
with the conditions
\begin{equation}\label{eq:normalization1}
x+\sum_{i=0}^{q-1}y_i=z
\end{equation}
\begin{equation}\label{eq:normalization2}
Q=\ee^x+\sum_{i=0}^{q-1}\ee^{y_i}
\end{equation}
and
\begin{equation}\label{eq:normalization3}
P=\frac{n^0}{Q}+W=\frac{n^0}{Q}+\frac{1-n^0}{(\ee^h-1)\ee^x+Q}.
\end{equation}

Solving equation~\eqref{eq:normalization3} with respect $ \ee^x $ , calling 
$\Gamma=PQ=n^0+WQ$ we get:
\begin{equation}\label{eq:solutionp1}
\frac{\ee^x}{Q}=\frac{1-\Gamma}{\Gamma-n^0}\frac{1}{\ee^h-1}
\end{equation}
and thus, plugging it in the first equation of~\eqref{eq:focforcing}, we
obtain:
\begin{equation}\label{eq:solutionp2}
x=\frac{z}{\ee^h-1}\frac{1-\Gamma}{\Gamma-n^0}\left[(1-\ee^h)n^0+\ee^h \Gamma 
\right]
\end{equation}

Plugging the previous result in~\eqref{eq:solutionp1}, we get Q.
On the other hand form equation 1 in~\eqref{eq:focforcing} we obtain 
\begin{equation}\label{eq:solutionp3}
\frac{1-n^0}{(\ee^h-1)\ee^x+Q}=P-\frac{n^0}{Q}
\end{equation}

plugging everything in equation~\eqref{eq:focforcing} we obtain the following 
system 
of equations written in term of $\Gamma$ :
\begin{equation}\label{eq:focforcinggamma}
\begin{cases}
x\ee^{-x}=\frac{z}{\ee^h-1} \left[(1-\ee^h)n^0+\ee^h \Gamma 
\right]\frac{1-\Gamma}{\Gamma-n^0} 
\ee^{-\frac{z}{\ee^h-1}\frac{1-\Gamma}{\Gamma-n^0}\left[(1-\ee^h)n^0+\ee^h 
\Gamma \right]} \\ 
y\ee^{-y}=\frac{z}{\ee^h-1} \Gamma\frac{1-\Gamma}{\Gamma-n^0} 
\ee^{-\frac{z}{\ee^h-1}\frac{1-\Gamma}{\Gamma-n^0}\left[(1-\ee^h ) n^0+\ee^h 
\Gamma \right]} & \text{q-1 times}
\end{cases}
\end{equation}

From previous equation~\eqref{eq:focforcinggamma} we can infer the structure of 
the solutions.

Both equations have the same shape:
\begin{equation}\label{eq:focforcingshape}
x\ee^{-x}= c
\end{equation}
where c is a constant to be determined auto-consistently.

If $c$ is negative  the equation has only one negative (thus unphysical) 
solution, if c is greater than $\ee^{-1}$ 
it has no solution otherwise it admits solution which can be expressed in 
terms of Lambert W functions~\cite{Corless1996}:
\begin{equation}\label{eq:xlam1}
x_-=-\fn{W}_0(-c)
 \end{equation} 
and
 \begin{equation}\label{eq:xlam2}
x_+=-\fn{W}_{-1}(-c)
 \end{equation} 
 where $\fn{W}_0,\fn{W}_{-1}$ represents the two real branches of Lambert W
function (using the notation of~\cite{Corless1996}). 
 It is trivial to check $x_-<1$ whereas $x_+>1$ and that for small values of 
$c$, $x_+>z$  and thus is to be discarded.
 
In our case $c$ is a complicate function of $\Gamma$ but, as above, once 
$\Gamma$ (and $z$)
is fixed we know that x and y can take only two values $x_{\pm}$ and $y_{\pm}$ 
as defined above and thus we can label
all the solutions using two integers $\alpha$ which counts the number of $x$ in 
+ state (which of course is either 0 or 1) 
and $L_+$ which counts the number of y in + state. 

If we define, for notational ease: 
\begin{equation}\label{eq:defB}
B(\Gamma)=\frac{z}{\ee^h-1} \Gamma\frac{1-\Gamma}{\Gamma-n^0} 
\ee^{-\frac{z}{\ee^h-1}\frac{1-\Gamma}{\Gamma-n^0}\left[(1-\ee^h ) n^0+\ee^h 
\Gamma \right]}
\end{equation}

and
\begin{equation}\label{eq:defA}
A(h,\Gamma)= \left[(1-\ee^h)n^0+\ee^h \Gamma \right]
\end{equation}
it is easy to check that solutions have the following hierarchy in the admissible range 
$\Gamma \in [n^0,1]$:
\begin{equation}\label{eq:hierarchy}
y_-(A(0,\Gamma)B(\Gamma))\leq x_-(A(h,\Gamma)B(\Gamma)) \leq 1 \leq 
x_+(A(h,\Gamma)B(\Gamma))\leq y_+(A(0,\Gamma)B(\Gamma))
\end{equation}

The normalization equation~\eqref{eq:normalization1} will then become, given 
integers $\alpha$ and $L_+$:
\begin{equation}\label{eq:norm_consistency}
\begin{split}
\alpha & x_+\bigg( \frac{z}{\ee^h-1} \left[(1-\ee^h )n^0+\ee^h \Gamma 
\right]\frac{1-\Gamma}{\Gamma-n^0} 
\ee^{-\frac{z}{\ee^h-1}\frac{1-\Gamma}{\Gamma-n^0}\left[(1-\ee^h)n^0+\ee^h 
\Gamma \right]}\bigg) \\ 
+ & (1-\alpha ) x_-\left(\frac{z}{\ee^h-1} \left[(1-\ee^h)n^0+\ee^h \Gamma 
\right]\frac{1-\Gamma}{\Gamma-n^0} 
\ee^{-\frac{z}{\ee^h-1}\frac{1-\Gamma}{\Gamma-n^0}\left[(1-\ee^h)n^0+\ee^h 
\Gamma \right]}\right)+ \\ 
L_+ & y_+\bigg(\frac{z}{\ee^h-1} \Gamma\frac{1-\Gamma}{\Gamma-n^0} 
\ee^{-\frac{z}{\ee^h-1}\frac{1-\Gamma}{\Gamma-n^0}\left[(1-\ee^h ) n^0+\ee^h 
\Gamma \right]}\bigg) \\ 
+ & (q-1-L_+)y_-\bigg(\frac{z}{\ee^h-1} \Gamma\frac{1-\Gamma}{\Gamma-n^0} 
\ee^{-\frac{z}{\ee^h-1}\frac{1-\Gamma}{\Gamma-n^0}\left[(1-\ee^h ) n^0+\ee^h 
\Gamma \right]}\bigg)=z.
\end{split}
\end{equation}
Its solutions will give the $\Gamma$ auto-consistently.

Since the previous substitution is valid only when ~\eqref{eq:solutionp2} is 
assumed we can rewrite the equation as:
\begin{equation}\label{eq:norm_consistency2}
\boxed{
\begin{split}
L_+ & y_+\bigg(\frac{z}{\ee^h-1} \Gamma\frac{1-\Gamma}{\Gamma-n^0} 
\ee^{-\frac{z}{\ee^h-1}\frac{1-\Gamma}{\Gamma-n^0}\left[(1-\ee^h ) n^0+\ee^h 
\Gamma \right]}\bigg) \\ 
+ & (q-1-L_+)y_-\bigg(\frac{z}{\ee^h-1} \Gamma\frac{1-\Gamma}{\Gamma-n^0} 
\ee^{-\frac{z}{\ee^h-1}\frac{1-\Gamma}{\Gamma-n^0}\left[(1-\ee^h ) n^0+\ee^h 
\Gamma \right]}\bigg)\\ 
=&z-\frac{z}{\ee^h-1}\frac{1-\Gamma}{\Gamma-n^0}\left[(1-\ee^h)n^0+\ee^h \Gamma 
\right]
\end{split}}
\end{equation}

\subsubsection{Stability }

When $n^\alpha=1-n^0$ all the values $n_a^\alpha=0$ for $\alpha>1$ because of 
the normalization constraints; the Lagrangian function, thus, becomes ( here 
$z=2\frac{\eta}{\lambda}$):
\begin{equation}\label{eq:lagrangian1f}
\begin{split}
\fn{L}(\bvec{n};z,\bvec{h})&=(\beta_{1}-1)\left(n^{1}+\sum_{a} 
n_a^{1}\right)-(\zeta_0-1) (n^0-\sum_a n_a^0)- h_{1} n_1^{1}\\+&\sum_a n_a^0 
\log (n_a^0)+ n_a^{1}\log(n_a^{1})-\frac{z}{2}(n_a)^2 
\end{split}
\end{equation}

To check the stability and the nature of these stationary points we have to 
check the Hessian $\mat{L}$ of the Lagrangian restricted to the tangent space 
$\sset{T}$ to the constraints manifold in the stationary point.

The stationary point will be a (local minimum) if and only if:
\begin{equation}\label{eq:SOC}
\mathbf{y}^T\mat{L}\mathbf{y}>0 \text{ for any } \mathbf{y} \in \sset{T}
\end{equation}
In our case the constraint are linear; therefore the tangent space $\sset{T}$ is 
a $2q-2$ dimensional space and it can be easily seen to be spanned by the 
orthonormal base:
\begin{equation}\label{eq:tangent_space}
(\bvec{e}_i)_j=\frac{1}{\sqrt{2}}(\delta_{ij}-2\delta_{2q+1+j\pmod{2},j}) 
\end{equation}
where $i \in \{1,..,2q\}$ and $j\in \{1,..,2q+2\}$.
The projection operator then is given by the matrix
\begin{equation}\label{eq:projection}
\mat{M}_{ij}=(\bvec{e}_i)_j 
\end{equation}
Any vector $\bvec{y}$ of $\sset{T}$ can be expressed by a general vector 
$\bvec{v}$ of $\mathbb{R}^{2q}$ as $\bvec{y}=\mat{M}\bvec{v}$. 
Equation~\eqref{eq:SOC}
can be then expressed as:
\begin{equation}\label{eq:SOC_2}
\bvec{v}^{T}\mat{M}^{T}\mat{L}\mat{M}\bvec{v}>0
\end{equation}
Thus in order to check the stability of a stationary point in the constrained 
problem we can simply  apply the usual Hessian criteria to the ``effective 
Hessian $\mat{H}_{\mathrm{eff}}=\mat{M}^{T}\mat{L}\mat{M}$

\subsection{Asymptotic Expansion of the solutions of FOC for $z\to 
\infty$}
\label{ap:asymptotic_expansion}

In the case of $z\to \infty$ case it is easy to see that the minimum of 
$\fn{F}(\bvec{n};\frac{\eta}{\lambda},\bvec{h})$ must correspond to states 
$\bvec{n}^{*}_{a}$ corresponding to $ n_a=1$ and $n_b=0$ for $a=1,\ldots,q$ and  $\forall\ \! 
b \neq a$.   

In order to gain some insight on which of these solutions is the true minimum of 
the free energy for large but finite $z$ we have to make and asymptotic 
expansion around $z=\infty$.

An asymptotic expansion of $n_a(z)$ can be derived from equation~\eqref{eq:foc_def} for large 
$z$. It is easy to verify that the leading correction is extremely small, i.e. 
\begin{equation}\label{eq:asymform}
n_b(z)\simeq\delta_{a b}+O( \ee^{-z})
\end{equation}
The detailed calculation of the leading order correction is carried out in Ref. \cite{DeLuca}. Here we remark that {\em 
i)} given the size of the correction, the asymptotic limit is representative also of the regime where $z$ is only 
moderately large. {\em ii)} the free energy for $z$ large is approximately given by
\begin{equation}
 \fn{F}\simeq -\frac{z}{2} + \sum_{b=0}^q n^b\log(n^b)-n^a 
h_a +z\ee^{-z}\big(-n^a h_a\ee^{h_a}+\sum_{b\neq a} n^b h_b\ee^{h_b} 
\big)+ o(\ee^{-z})  .
\end{equation}
For this expression it is clear that the global minimum for large $z$ is the solution with the maximum  $n^a h_a$. In 
the case in which two or more direction have the same value of $h_i n^i$ the solution 
with the bigger value of $h_i$ will prevail. 

\subsection{Tradeoff}
\label{sec:tradeoff}

Let us now discuss the generalization of the above results to cases where informed individuals might be \emph{less 
social} then their co-specific uniformed fellows. Several works, for example \cite{Couzin}, have assumed that there is 
a 
trade-off between sociality and the ability of individuals to store, gather or process information. There is no 
conclusive evidence that such trade-offs exist, to the best of our knowledge. Evidences are mounting that 
individual-level information and social processes can be both present in collective decision processes in fish 
groups~\cite{Miller2013, Herbert-Read2013}. However, since this aspect has been included as an important 
ingredient in other models, it is important to explore its reelvance in the present context.

Our theoretical framework can be extended to account for this aspect in two different ways: either by making informed 
individuals promote the formation of links at a lower rate or by generalizing the choice behavior to a 
stochastic probabilistic model. 
Here we show that the qualitative results 
discussed in the main paper are kept significantly unchanged in both cases.

\subsubsection{Heterogeneous link formation rates}

A natural way to introduce a tradeoff between sociality and information is to 
assume that  \emph{informed individuals shall promote link creation at a 
lower rate}.

Under this assumption it is possible to calculate the exact invariant 
distribution and to proceed with the same calculations of the simpler case 
discussed in the main article. In particular the large $z$ and small $z$ 
solutions are the same. Here we report the main results, without repeating lengthy 
derivations. 

We assume that individuals that are informed about direction $\alpha$ 
shall promote link creation with rate $\eta^\alpha$ whereas uninformed 
individuals shall promote link creation with rate $\eta^0$. While we keep the dependence on $\alpha$ in $\eta$, it is 
reasonable to assume that $\eta^\alpha=\eta^1$ takes the same value, irrespective of the preferred direction $\alpha\neq 
0$, for all informed individuals. 

In particular we have that the link creation rate for the creation of a link 
between nodes $i$ and $j$  is :
 \begin{enumerate}
    \item $\frac{\eta^{\alpha_i}+\eta^{\alpha_j}}{N-1}$ if $a_j=a_j$ (here $\alpha_i=0$  denotes uninformed individuals)
\item $0$ otherwise
 \end{enumerate}

The invariant measure therefore reads

 \begin{equation}\label{eq:invariantmeasure_tradoff}
\pi(\omega)=\frac{1}{\sset{Z}}
\ee^{\sum_i h_{\alpha_i} 
\delta_{a_i {\alpha_i}}}\prod_{j<i}\left(\frac{\delta_{a_ia_j}(\eta^{\alpha_i}
+\eta^{\alpha_j}) } { \lambda(N-1) } \right)^{g_{ij}} .
\end{equation}

In order to obtain the distribution in term of population we have to distinguish between 
two different types of links.
Let us define $M^{\alpha \beta}_a$ denote the number of links between individuals  that are in state $a$ but 
would prefer to be in state $\alpha$ and  individuals that are in state $a$ but  would prefer to be in state 
$\beta$.For simplicity we shall extend the notation described above denoting with preference $0$ the individuals 
with no preference. Let $N_a^\alpha$ denote the number of individuals 
that are in state $a$ but would prefer to be in state $\alpha$.

It is trivial to infer from eq.~\eqref{eq:invariantmeasure_tradoff}  that, in term of these 
quantities, the invariant distribution reads
\begin{equation}\begin{split}
 \pi(\bvec{N} & ,\bvec{M})=\frac{1}{\mathcal{Z}}\frac{N!}{
\prod_a N_a^0!\prod_{a\alpha  
}N_a^{\alpha}!}\ee^{
\sum_a h_a N_a^a } 
 \prod_{a\in 
S, \alpha \in S\cup\{0\}}\binom{\frac{N^{\alpha}_a(N^\alpha_a-1)}{2}}{M^{\alpha 
\alpha}_a}\left(\frac{2 
\eta^\alpha}{\lambda(N-1)}\right)^{M^{\alpha \alpha}_{a}} \\ &\prod_{\beta \in S\cup\{0\}, \beta \neq \alpha} 
\binom{N^\alpha_a N^\beta_a}{M^{\alpha\beta}_a} \left(\frac{ 
\eta^\alpha+\eta^\beta}{\lambda(N-1)}\right)^{M^m_{a}}.
\end{split}
\end{equation}

Summing over $M$'s and taking the logarithm we get that
\begin{equation}
 \log(\pi)=\sum_{a\alpha} N^\alpha_a \log(N^\alpha_a)-\sum_a h_a N_a^a -\left(\sum_{a \alpha \beta}
N^\alpha_a N^\beta_a \frac{\eta_\alpha +\eta_\beta}{\lambda(N-1)}\right) -\sum_{a \alpha 
}N^\alpha\frac{2\eta_\alpha}{\lambda(N-1)}
\end{equation}
thus leading to a free energy in the thermodynamic limit:

\begin{equation}
\label{eq:Fheterogeta}
F(\bvec{n}; z^\alpha, h_\alpha)=\sum_{a 
\alpha}n^\alpha_a\log(n_a^\alpha)-\sum_a h_\alpha n_a^a 
-\frac{1}{2}\sum_a\left(\sum_\alpha z^\alpha n_a^\alpha\right)n_a
\end{equation}
The saddle point equations can be derived in a straightforward manner, following the steps outlined above for the 
homogeneous case. 

The numerical solution of these equations can be studied as a function of the parameters. In particular, in the extreme 
case where $\eta^\alpha=0$ for all informed individuals ($\alpha>0$) exhibits the same hysteric behavior of the 
homogeneous case discussed above. The limit $z^0=\eta^0/\lambda\gg 1$ can also be studied in exactly the same way as 
above, with the same conclusion, that corrections to the coordinated solutions $n_b^*=\delta_{a b}$ are exponentially 
small in $z_0$. From Eq. (\ref{eq:Fheterogeta}) it is clear that again the dominant solution for $z^0$ large is the one 
with the largest product $n^ah_a$, exactly as in the homogeneous case. This conclusion also extends to the case where 
$z^\alpha=z^1>0$ for $\alpha\neq 0$.

\subsubsection{Generalized internal update rule}

A different way to introduce a tradeoff between sociality and preferences is to 
modify the update rule~\eqref{eq:conform}, by assuming that individuals weight their preferred choice and the choice 
taken by neighbors when updating their choices. One way to do this is to assume that 
 individual $i$, when an internal state update event occurs, will pick up randomly an internal state with a probability 
proportional to:
\begin{equation}\label{eq:update_rule}
P_i(a)\propto \ee^{h_{\alpha_i}\delta_{a\alpha_i}+\beta_i \sum_{j} g_{ij}\delta_{a a_j}}.
\end{equation}

The parameter$\beta_i$ measures to what extent individual $i$  takes into account the choices of the social group in 
making his decision. The effect of this choice is mostly evident by discussing the limiting cases:
\begin{enumerate}
 \item if $\beta_i\to 0 $,  individual $i$ will make its choices selfishly, without considering the choices 
of its neighbouring individuals;
\item if $\beta_i \to \infty$, individual $i$ will make its choices always conforming to its neighbourhood, 
if ths is not empty. Only when his neighborhood is empty, his choice will reflect the pieces of information it 
possesses: this is exactly the case discussed in the main paper.

\end{enumerate}

\begin{figure}
\includegraphics[width=\textwidth]{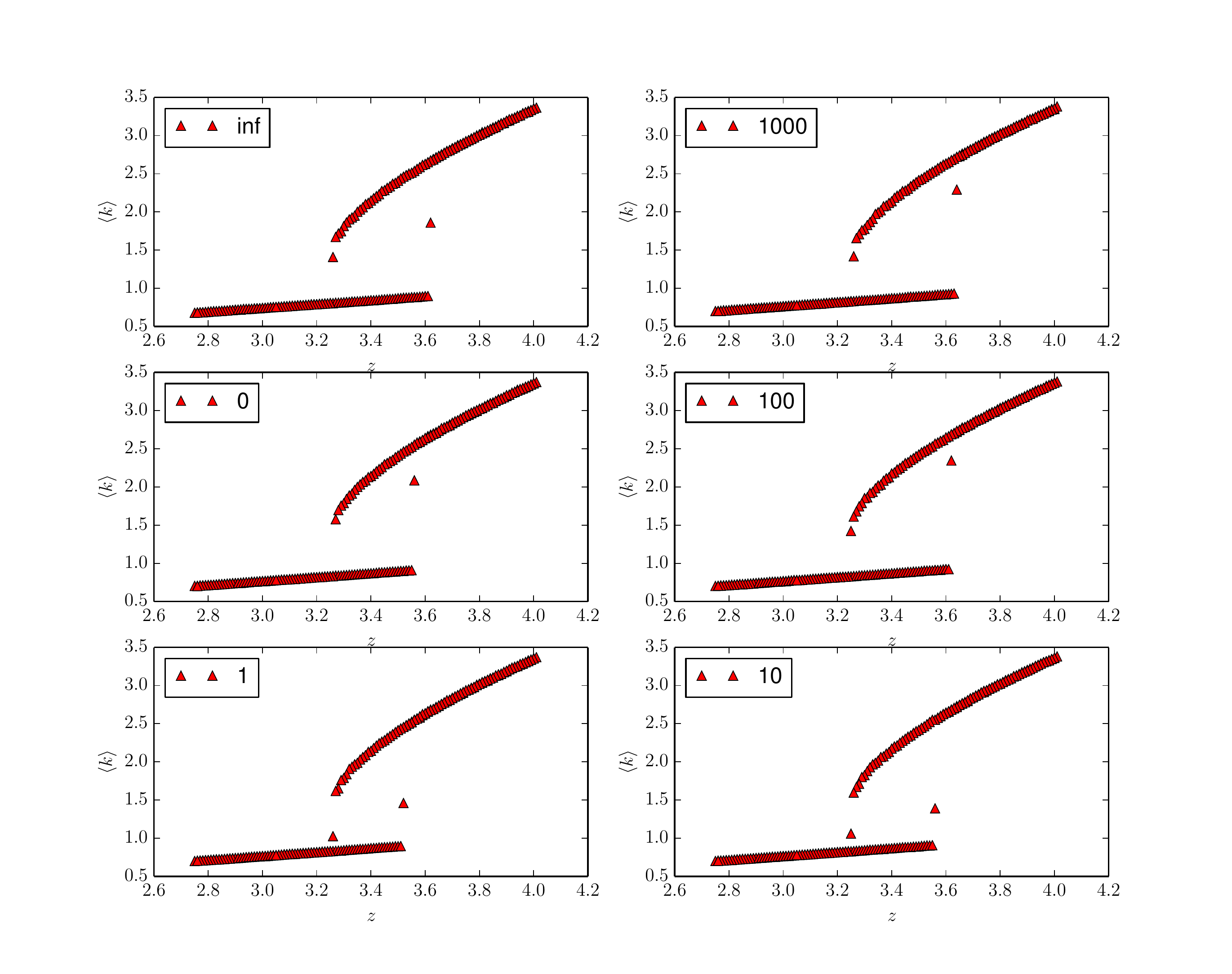}\label{fig:beta}
\caption{Plot of the average degree  obtained in numerical simulations of the model with stochastic choice update. These 
are based on 100 runs, with $N=5000$ 
nodes and parameter $n^0=0.99$, $n^1=0.01$ and $h_1=0.05$ for different values of $\beta$ (the value is written in the 
legend) for the informed individuals (the uninformed ones have $\beta_i=\infty$).}
\end{figure}
As one can see, $\beta_i$ interpolates between a fully pro-social behaviour and a completely individualistic one; the 
tradeoff may be simply introduced assuming that uniformed individuals have $\beta_i=\infty$ whereas informed 
individuals have lower $\beta_i$.

The drawback of this extension of the model is that, as soon as $\beta_i<\infty$, the system is not analytically 
solvable and can only be investigated by numerical stochastic simulations. Fig. \ref{fig:beta} reports a series of 
numerical simulations of this extended model. This shows that introducing a trade-off between sociality and preference 
does not qualitatively change 
the behaviour of the system; the only appreciable difference being the fact that the symmetric solution 
becomes unstable at  lower values of $z$. Therefore, even though quantitative results differ, the qualitative picture 
described in the article holds true also in this more general model.


\begin{thebibliography}{10}

\bibitem{Sumpter2010}
Sumpter DJT.
\newblock {Collective Animal Behavior}.
\newblock Princeton University Press. Princeton University Press; 2010.
\newblock Available from: \url{http://books.google.it/books?id=nB67IvU1qlAC}.

\bibitem{krause2002living}
Krause J, Ruxton GD.
\newblock {Living in groups}.
\newblock Oxford University Press; 2002.

\bibitem{Krause2010}
Krause J, Ruxton GD, Krause S.
\newblock {Swarm intelligence in animals and humans.}
\newblock Trends in ecology \& evolution. 2010 Jan;25(1):28--34.
\newblock Available from: \url{http://www.ncbi.nlm.nih.gov/pubmed/19735961}.

\bibitem{Surowiecki2005}
Surowiecki J.
\newblock {The Wisdom of Crowds}.
\newblock Anchor; 2005.
\newblock Available from: \url{http://www.worldcat.org/isbn/0385721706}.

\bibitem{Lachlan1998}
Lachlan R, Crooks L, Laland K.
\newblock {Who follows whom? Shoaling preferences and social learning of
  foraging information in guppies.}
\newblock Animal behaviour. 1998 Jul;56(1):181--90.
\newblock Available from: \url{http://www.ncbi.nlm.nih.gov/pubmed/9710476}.

\bibitem{Conradt2011}
Conradt L.
\newblock {Models in animal collective decision-making: information uncertainty
  and conflicting preferences}.
\newblock Interface Focus. 2011 Dec;2(2):226--240.
\newblock Available from:
  \url{http://rsfs.royalsocietypublishing.org/cgi/doi/10.1098/rsfs.2011.0090}.

\bibitem{sumpter2008consensus}
Sumpter DJT, Krause J, James R, Couzin ID, Ward AJW.
\newblock {Consensus decision making by fish}.
\newblock Current Biology. 2008;18(22):1773--1777.

\bibitem{Couzin2011}
Couzin ID, Ioannou CC, Demirel G, Gross T, Torney CJ, Hartnett A, et~al.
\newblock {Uninformed individuals promote democratic consensus in animal
  groups.}
\newblock Science (New York, NY). 2011 Dec;334(6062):1578--80.
\newblock Available from: \url{http://www.ncbi.nlm.nih.gov/pubmed/22174256}.

\bibitem{Ward2011}
Ward AJW, Herbert-Read JE, Sumpter DJT, Krause J.
\newblock {Fast and accurate decisions through collective vigilance in fish
  shoals.}
\newblock Proceedings of the National Academy of Sciences of the United States
  of America. 2011 Feb;108(6):2312--5.
\newblock Available from:
  \url{http://www.pubmedcentral.nih.gov/articlerender.fcgi?artid=3038776\&tool%
=pmcentrez\&rendertype=abstract}.

\bibitem{Miller2013}
Miller N, Garnier S, Hartnett AT, Couzin ID.
\newblock {Both information and social cohesion determine collective decisions
  in animal groups.}
\newblock Proceedings of the National Academy of Sciences of the United States
  of America. 2013 Feb;110(13).
\newblock Available from: \url{http://www.ncbi.nlm.nih.gov/pubmed/23440218}.

\bibitem{Conradt2011a}
Conradt L.
\newblock {Collective behaviour: When it pays to share decisions.}
\newblock Nature. 2011 Mar;471(7336):40--1.
\newblock Available from: \url{http://www.ncbi.nlm.nih.gov/pubmed/21368814}.

\bibitem{berdahl2013emergent}
Berdahl A, Torney CJ, Ioannou CC, Faria JJ, Couzin ID.
\newblock {Emergent Sensing of Complex Environments by Mobile Animal Groups}.
\newblock Science. 2013;339(6119):574--576.

\bibitem{Brown2003}
Brown C, Laland KN.
\newblock {Social learning in fishes: a review}.
\newblock Fish and Fisheries. 2003 Sep;4(3):280--288.
\newblock Available from:
  \url{ http://doi.wiley.com/10.1046/j.1467-2979.2003.00122.x}.

\bibitem{Couzin2005}
Couzin ID, Krause J, Franks NR, Levin Sa.
\newblock {Effective leadership and decision-making in animal groups on the
  move.}
\newblock Nature. 2005 Feb;433(7025):513--6.
\newblock Available from: \url{http://www.ncbi.nlm.nih.gov/pubmed/15690039}.

\bibitem{Guttal2010}
Guttal V, Couzin ID.
\newblock {Social interactions, information use, and the evolution of
  collective migration.}
\newblock Proceedings of the National Academy of Sciences of the United States
  of America. 2010 Sep;107(37):16172--7.
\newblock Available from:
  \url{http://www.pubmedcentral.nih.gov/articlerender.fcgi?artid=2941337\&tool%
=pmcentrez\&rendertype=abstract}.

\bibitem{Guttal2011}
Guttal V, Couzin ID.
\newblock {Leadership, collective motion and the evolution of migratory
  strategies.}
\newblock Communicative \& integrative biology. 2011 May;4(3):294--8.
\newblock Available from:
  \url{http://www.pubmedcentral.nih.gov/articlerender.fcgi?artid=3187890\&tool%
=pmcentrez\&rendertype=abstract}.

\bibitem{Katz2011}
Katz Y, Tunstr\o~m Kr, Ioannou CC, Huepe C, Couzin ID.
\newblock {Inferring the structure and dynamics of interactions in schooling
  fish.}
\newblock Proceedings of the National Academy of Sciences of the United States
  of America. 2011 Nov;108(46):18720--5.
\newblock Available from:
  \url{http://www.pubmedcentral.nih.gov/articlerender.fcgi?artid=3219116\&tool%
=pmcentrez\&rendertype=abstract}.

\bibitem{Vicsek2012}
Vicsek T, Zafeiris A.
\newblock {Collective motion}.
\newblock Physics Reports. 2012 Mar;p. 71--140.
\newblock Available from: \url{
  http://linkinghub.elsevier.com/retrieve/pii/S0370157312000968}.

\bibitem{Chate2008}
Chat\'{e} H, Ginelli F, Gr\'{e}goire G, Peruani F, Raynaud F.
\newblock {Modeling collective motion: variations on the Vicsek model}.
\newblock The European Physical Journal B-Condensed Matter and Complex Systems.
  2008;64(3):451--456.

\bibitem{Huepe2011}
Huepe C, Zschaler G, Do AL, Gross T.
\newblock {Adaptive-network models of swarm dynamics}.
\newblock New Journal of Physics. 2011 Jul;13(7):073022.
\newblock Available from:
  \url{http://stacks.iop.org/1367-2630/13/i=7/a=073022?key=crossref.49dadee94c%
09c539906fc49c3d1d1b5c}.

\bibitem{Gross2008}
Gross T, Blasius B.
\newblock {Adaptive coevolutionary networks: a review.}
\newblock Journal of the Royal Society, Interface / the Royal Society. 2008
  Mar;5(20):259--71.
\newblock Available from:
  \url{http://www.pubmedcentral.nih.gov/articlerender.fcgi?artid=2405905\&tool%
=pmcentrez\&rendertype=abstract}.

\bibitem{Marsili2010}
Marsili M, Vega-Redondo F, Ehrhardt G.
\newblock {Networks emerging in a volatile world}.
\newblock EUI Working Paper ECO. 2008;.

\bibitem{Ehrhardt2006}
Ehrhardt G, Marsili M, Vega-Redondo F.
\newblock {Phenomenological models of socioeconomic network dynamics}.
\newblock Physical Review E. 2006 Sep;74(3):36106.
\newblock Available from:
  \url{http://link.aps.org/doi/10.1103/PhysRevE.74.036106}.

  \bibitem{Krause1992}
Krause, J., Bumann, D., \& Todt, D., 1992
{Relationship between the position preference and nutritional state of 
individuals in schools of juvenile roach (Rutilus rutilus).}
\newblock \emph{Behavioral Ecology and Sociobiology} \textbf{30} (3-4), 177–180. 
\newblock Available from:  \url{http://dx.doi.org/10.1007/BF00166700}).


\bibitem{Petitgas2010}
Petitgas P, Secor DH, McQuinn I, Huse G, Lo N.
\newblock {Stock collapses and their recovery: mechanisms that establish and
  maintain life-cycle closure in space and time}.
\newblock ICES Journal of Marine Science. 2010 Jun;67(9):1841--1848.
\newblock Available from:
  \url{http://icesjms.oxfordjournals.org/cgi/doi/10.1093/icesjms/fsq082}.

\bibitem{Huse2002}
Huse G, Railsback S, Fern\"{o} A.
\newblock {Modelling changes in migration pattern of herring: collective
  behaviour and numerical domination}.
\newblock Journal of Fish Biology. 2002 Mar;60(3):571--582.
\newblock Available from: \url{http://doi.wiley.com/10.1006/jfbi.2002.1874}.

\bibitem{Herbert-Read2013}
Herbert-Read JE, Krause S, Morrell LJ, Schaerf TM, Krause J, Ward aJW.
\newblock {The role of individuality in collective group movement.}
\newblock Proceedings Biological sciences / The Royal Society. 2013
  Feb;280(1752):20122564.
\newblock Available from: \url{http://www.ncbi.nlm.nih.gov/pubmed/23222452}.

\bibitem{ICES2011}
ICES.
\newblock {Report of the herring assessment working group for the area south of
  62 deg N}.
\newblock ICES; 2011.
\newblock Available from: \url{http://www.ncbi.nlm.nih.gov/pubmed/23222452}.


\bibitem{ICES2012}
ICES.
\newblock {ICES Advices. Book 6.}
\newblock Copenhagen: ICES; 2012.

\bibitem{ICCAT2012}
ICCAT.
\newblock {Report of the 2012 Atlantic Blufin Tuna stock assesment session}.
\newblock Madrid: ICCAT; 2012.
\newblock Available from:
  \url{http://www.iccat.int/Documents/Meetings/SCRS2012/SCI-003\_ENG.pdf}.



\end{thebibliography}

\begin{thebibliography}{1}

\bibitem{Corless1996}
R.~M. Corless, G.~H. Gonnet, D.~E.~G. Hare, D.~J. Jeffrey, and D.~E. Knuth.
\newblock {\em On the Lambert W function}.
\newblock { Advances in Computational Mathematics}, 5(1):329--359, December
  1996.

\bibitem{Norris1998}
J.~R. Norris.
\newblock {\em {Markov Chains}}.
\newblock Cambridge University Press, Cambridge, 1998.

\bibitem{DeLuca}
G. De~Luca
\newblock {\em Decision making in complex environments: an adaptive network 
approach} 
\newblock Ph.D. Thesis, Trieste, 2013 \url{http://hdl.handle.net/1963/7203} .

\bibitem{Couzin} 
V.~Guttal , I.~D. Couzin. 
\newblock {\em Social interactions, information use, and the evolution of collective migration}. 
\newblock Proceedings of the National Academy of Sciences of the United States of America. 2010;{\bf 107} (37): 16172-7.

\bibitem{Miller2013}
Miller N, Garnier S, Hartnett AT, Couzin ID.
\newblock {Both information and social cohesion determine collective decisions
  in animal groups.}
\newblock Proceedings of the National Academy of Sciences of the United States
  of America. 2013 Feb;110(13).
\newblock Available from: \url{http://www.ncbi.nlm.nih.gov/pubmed/23440218}.

\bibitem{Herbert-Read2013}
Herbert-Read JE, Krause S, Morrell LJ, Schaerf TM, Krause J, Ward aJW.
\newblock {The role of individuality in collective group movement.}
\newblock Proceedings Biological sciences / The Royal Society. 2013
  Feb;280(1752):20122564.
\newblock Available from: \url{http://www.ncbi.nlm.nih.gov/pubmed/23222452}.

\end{thebibliography}
\end{document}